\documentclass[letterpaper]{article}

\usepackage[T1]{fontenc}

\usepackage{geometry}
\usepackage{setspace}

\usepackage[numbers,sort&compress]{natbib}

\usepackage{graphicx}
\usepackage{float}
\usepackage{hyperref}
\usepackage{listings}
\usepackage{xcolor}
\usepackage{subcaption}
\newfloat{scheme}{htbp}{los}
\floatname{scheme}{Scheme}
\floatname{chart}{Chart}
\newfloat{graph}{htbp}{loh}

\usepackage{chemformula} 
\usepackage[version = 4]{mhchem} 

\usepackage{authblk}
\author[1]{William Dawson}
\author[2]{Louis Beal}
\author[3]{Marco Zaccaria}
\author[3]{Luigi Genovese}
\affil[1]{RIKEN Center for Computational Science, Kobe, Hyogo 650-0047, Japan}
\affil[2]{INRIA, France}
\affil[3]{University Grenoble Alpes, MEM, L\_Sim, F-38000 Grenoble, France}

\title{Coupling Quantum Mechanical Modeling and Molecular Dynamics on Heterogeneous Supercomputers for Studying Distal Mutation Effects on Drug Binding in HIV-1}
\date{*Email: william.dawson@riken.jp}

\begin{document}

\maketitle

\begin{abstract}
Predicting how protein mutations affect drug binding remains a major challenge, particularly when the mutations are distal from the binding site. In this study, we introduce a coupled simulation workflow that combines long-time-scale molecular dynamics (MD) with high-throughput quantum mechanical (QM) analysis to reveal the electronic structure signatures of mutation induced drug resistance in the HIV-1 protease. Our workflow leverages GPU-accelerated MD to generate conformational ensembles, and performs in-operando linear-scaling density functional theory (DFT) calculations on selected frames parallelized on a coupled partition of CPU nodes. This design enables efficient, massively parallel quantum analysis of protein-ligand complexes at atomic resolution. Using this approach, we investigate resistance to the antiviral Darunavir in a multi-mutant HIV-1 protease variant. By mapping the network of electronic interactions across the binding interface, our results highlight the critical role of conformational sampling and quantum insight in understanding distal mutation effects, and demonstrate a scalable computational strategy for studying complex biophysical mechanisms of drug resistance. 
We argue that such kind of analysis may pave the way for designing inhibitors that maintain binding stability against systemic, mutation-induced destabilization.
\end{abstract}

\section*{Keywords}

Electronic Structure, Molecular Dynamics, HIV, Drug Discovery, Supercomputing.

\section{Introduction}

In-silico studies of protein-ligand binding offer the promise of atomic level insight into binding trends and of predicting interaction strengths of novel molecules. Performing such studies requires the combination of multiple simulation techniques, each with its own strengths in certain aspects of chemistry~\cite{brazdova2013atomistic}. Because protein-ligand binding is a dynamic process involving highly flexible moieties at finite temperature, molecular dynamics based on classical potentials (MD) remains a popular choice. However, since classical potentials rely on a ball-and-stick model of physics --- lacking the explicit modeling of electrons --- MD is limited in the accuracy and insight it can provide~\cite{Dawson2022}.  Hence, researchers often incorporate quantum mechanical (QM) methods, most popularly Kohn-Sham Density Functional Theory (DFT)~\cite{hohenberg-inhomogeneous-1964,kohn-self_consistent-1965}, into their research workflow.

Even within these select two modeling techniques, there is a wide variety of numerical methods and software implementations. Pushing the limits of size and sampling breadth requires the development of high-performance implementations that can run on cutting edge supercomputers. In the past decade, computers composed of Graphics Process Units (GPUs) have dominated the top supercomputers in the world. However, the challenge of porting methods to GPUs (which may grow due to the influence of deep learning~\cite{Shinde2025}) means that many methods are only available as CPU implementations. Thus, universities and computing centers continue to offer substantial computational resources to researchers in the form of CPU only machines. 

A potential response to this fractured software ecosystem is for computing centers to offer both GPU and CPU partitions to users.  In this work, we propose a combined MD + QM workflow, where a classical trajectory of protein-ligand binding generated on the GPU partition is analyzed in-operando by QM calculations running on the CPU partition. The bundling of these calculation steps together allows for an overall faster processing time of the workflow, as QM calculations and subsequent post processing are performed on intermediate snapshots of the trajectory as soon as data are available.

As a demonstration of this methodology, we will use our workflow to study drug resistance to the essential medicine Darunavir (DRV) caused by mutations of the HIV-1 protease (HIV-1 PR). A previous combined in-silico / in-vitro study analyzed a set of 11 HIV-1 mutations that decrease binding to Darunavir~\cite{Henes2019}. While three mutations near the active site showed a loss of inhibitory power, intriguingly a separate set of 8 far away (``distal'' mutations) also caused substantial binding loss. In this study, we will utilize linear-scaling DFT~\cite{Ratcliff2017} to explicitly model these distal mutations. Our combination of advanced DFT methods with long MD trajectories will prepare the way for fully in-silico studies of novel mutations that may require extensive sampling to escape the local conformation of the reference structure. 


\section{Choice of Case Study: Darunavir and HIV-1 Protease}
Darunavir (DRV) is an HIV-1 protease inhibitor designed to be genetically robust via its binding to conserved residues in the enzymatic pocket (Asp29 and Asp30). These residues are adjacent to the catalytic element Asp25, and cannot mutate without distorting its geometry and compromising the protease's ability to cleave viral polyproteins (See Ghosh et al.).
However, a growing body of experimental evidence demonstrates that resistance can also arise from mutations located far from the inhibitor binding pocket, whose effects cannot be rationalized by local structural changes alone. In this context, to acquire DRV resistance, the virus needs to accumulate 3-4 or more mutations from a specific set, as identified by IAS-USA; namely: V11I, V32I, L33F, I47V, I50V, I54L, I54M, T74P, L76V, I84V, L89V. Three of these, I50V, I84V, I47V, do occur in the enzymatic pocket, changing its shape and preventing DRV from binding. The remaining eight (V11I, V32I, L33F, I54L, I54M, T74P, L76V, L89V), however,  are distal and drive DRV resistance via permissive epistasis, offsetting the energetic cost of the proximal mutations. (Ozen et al., 2011; Fun et al., 2012).
A particularly compelling example is a combined experimental and computational study by the group of C. Schiffer~\cite{Henes2019} of the HIV‑1 PR inhibitor Darunavir.

In the work, the authors investigated a series of HIV‑1 PR variants containing up to eleven mutations and quantified their effects on Darunavir binding affinity. While variants containing only a small number of active--site substitutions showed a moderate reduction in inhibitory potency, the co-occurrence of distal mutations within the broader genetic background corresponded to a dramatic loss of affinity, spanning more than five orders of magnitude (with the $K_i$ changing from less than 5 pM to 759.2 nM). Importantly, these distal mutations did not directly disrupt the inhibitor -- protein contacts observed in crystallographic structures, nor did they introduce obvious steric clashes within the active site. In addition to loss in inhibitor potency, the authors measured the change in catalytic efficiency, which was defined as the ratio of the turnover rate ($K_{\text{cat}}$) to the Michaelis–Menten constant ($K_{M}$). The values of $\frac{K_{\text{cat}}}{K_{M}}$ for the wild type, 2 mutation variant, and 11 mutation variant were $17.1 \, \mu\text{M}^{-1} \, \text{s}^{-1}$, $3.4 \pm 0.2 \, \mu\text{M}^{-1} \, \text{s}^{-1}$, and $1.3 \pm 0.4 \, \mu\text{M}^{-1} \, \text{s}^{-1}$, respectively, showing a modest decrease in efficiency in contrast to the substantial drop in inhibitor potency.

The authors proposed that the distal mutations act through indirect and non-additive mechanisms characteristic of epistasis. These changes collectively reshape the conformational ensemble of the PR. Molecular dynamics simulations revealed altered flap dynamics and changes in the stability of the closed, inhibitor--bound state. These effects illustrate a form of long--range allostery, whereby sequence changes distant from the binding site modulate inhibitor binding by subtly shifting the free energy landscape rather than by modifying specific local interactions. Such mechanisms are inherently difficult to capture using static structural models. Classical molecular mechanics simulations are well suited to explore conformational flexibility over long timescales but rely on empirical force fields that do not explicitly represent electronic structure. As a result, they may struggle to accurately describe the energetic consequences of distal mutations, especially when these involve cooperative effects or subtle redistribution of charge density. These distal epistatic mutations cannot be linked to independent changes in the binding site geometry;  a methodological approach that goes beyond local and static interpretations of binding is therefore warranted.

The Darunavir resistance mutations examined by Henes et al. therefore represent an ideal test case for a combined MD+QM workflow. By combining extensive molecular dynamics sampling with linear--scaling density functional theory calculations, it becomes possible to explicitly assess how long--range sequence variations affect the electronic and energetic structure of the protein--ligand complex across relevant conformational states.
Our workflow enables the investigation of mutation--induced interaction patterns at the amino-acid resolution, while retaining the conformational diversity captured by long MD trajectories. In doing so, it lays the groundwork for fully in‑silico studies of resistance mechanisms that are inaccessible to purely classical or purely static methodologies.

\section{Combined Workflow Methodology}

\subsection{Preparation of the HIV-1 PR Dataset}

\begin{figure}
    \centering
    \includegraphics[width=0.5\textwidth]{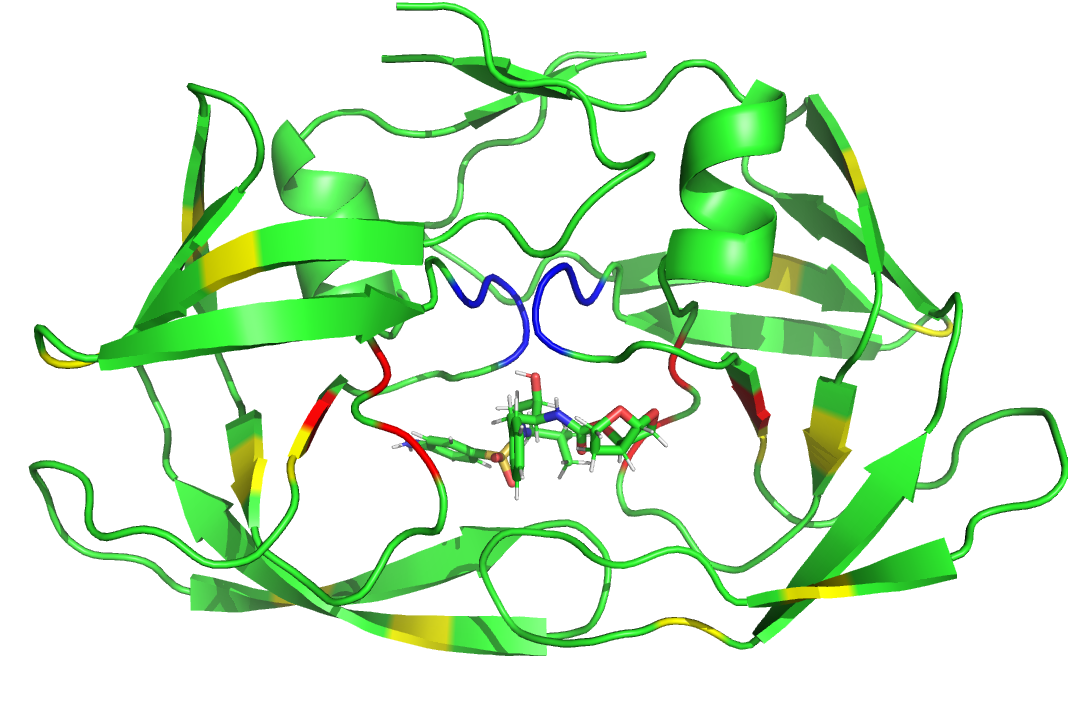}
    \caption{Mutation sites of the HIV-1 PR. The protein is represented in the cartoon style and the ligand with licorice. The blue residues are the catalytic triad (D25, T26 and G27), red residues nearby mutations, and yellow residues the distal mutations.}
    \label{fig:mut}
\end{figure}

We study three variants of HIV-1: wild type, 2 (V82F, I84V: active site) mutations, and 11 mutations: three in the active site: V32I, V82F, I84V; five proximal/flap: L33F, K45I, M46I, I54L, L76V; three distal: I13V, G16E, A71V. In the context of this work, and for uniformity with Hesen at al., we will define as distal all mutations not occurring within the active site, and dispense of the nuanced "proximal" category. These variants are based on three PDB structures: 6OPS (wild type), 6OPT (2 mutations), and 6OPZ (11 mutations)~\cite{Henes2019}. Each structure was prepared for MD simulation using the CHARMM-GUI solution builder tool~\cite{Jo2008,Brooks2009,Lee2016}, primarily with default settings. Crystallographic waters were excluded, and the system was solvated in an explicit 0.15 M NaCl solution. Cell sizes are set using an edge distance of 12~\AA~to correspond with the previous study. This results in box sizes of 79, 81, and 80~\AA~for the wild type, 2 mutation, and 11 mutation variants, respectively. We simulate HIV-1 PR in its dimer form. D25 of one chain of the catalytic triad is modeled in its deprotonated state~\cite{Smith1996}. We deprotonate the prime side residue consistent with neutron scattering findings~\cite{Gerlits2016}. A visualization of the mutated residues and the catalytic site is shown in Figure~\ref{fig:mut}.

\subsection{MD, QM, and Complexity Reduction}

For the MD portion of the workflow, we use the GENESIS software package~\cite{Jung2024}. Calculations are performed using the CHARMM36m forcefield~\cite{Huang2017} for the protein, the CHARMM General Forcefield (CGenFF)~\cite{Vanommeslaeghe2012,Vanommeslaeghe20122} for the ligand, and TIP3P~\cite{Jorgensen1983} for water. During the optimization stage, we perform steepest descent minimization for 40,000 steps. The equilibration stage performs 250,000 steps at 303.15K using the NVT thermostat, integrated with a 1~fs timestep and the velocity Verlet algorithm. The production phase runs 50,000,000 steps at 303.15K using the NPT thermostat at ambient pressure, integrated with a 2~fs timestep and the RESPA algorithm (80~ns of total simulation time). For all simulations we use the SHAKE algorithm to constrain bonds involving hydrogen and a cutoff distance of 12.0~\AA~. During the production runs, we write intermediate trajectories every 1,000,000 steps for QM postprocessing, resulting in 40 snapshots for each system. 

We utilize the BigDFT code for the QM portion of the calculation, which is a DFT code based on Daubechies wavelets~\cite{Ratcliff2020}. We extract the protein-ligand system and only the water molecules within 3~\AA~of any atom of the complex. This results in system sizes of approximately 5,200 atoms, a size that can be efficiently treated at a full QM level of theory by the linear scaling mode of BigDFT~\cite{Mohr2015}. Since neutralizing ions are not explicitly included in the QM calculation, we set the net charge based on the protonation state. We use the PBE functional~\cite{10.1103/PhysRevLett.77.3865}, a grid spacing of 0.45 Bohr, and HGH pseudopotentials~\cite{Willand2013} with Non-Linear Core Correction terms. 
While computing precise protein-ligand interaction energies may require a more advanced density functional~\cite{Chan2022}, as we focus on density based descriptors of interactions PBE will be sufficient~\cite{Dawson2023}.

MM calculations may be post processed with a large-scale QM calculation for a number of reasons. One possibility might be to compute a binding energy correction taking advantage of the improved accuracy DFT provides over classical forcefields~\cite{Ryde2016}. For our study, we are interested in the improved insight that can be provided by a quantum mechanical method~\cite{Dawson2022}. Post processing of the QM calculations is performed using the complexity reduction framework (QM-CR) we proposed in a series of previous publications\cite{Mohr2017,10.1021/acs.jctc.9b01152,Dawson2023}. From this analysis, we are able to first partition the target drug molecule into a set of chemically meaningful fragments. Then, the framework can generate interaction graphs between drug fragments and protein amino acids, classifying interactions as driven by short range chemical bonding or long range electrostatic interactions. The mathematical details of the QM-CR scheme relevant for this study are reviewed in the Appendix. 

This approach has been used in several applications of protein binding; here, we highlight two closely related studies. Recently, QM-CR was used to post process trajectories coming from molecular dynamics simulations of the SARS-CoV-2 main-protease bound to $\alpha$-ketamide inhibitors~\cite{Genovese2023}. In another, we used QM-CR to understand and predict mutations of the SARS-CoV-2 spike protein~\cite{Zaccaria2024}. We note that similar analysis is available when using the Fragment Molecular Orbital method~\cite{Fedorov2007}, and its combination with classical molecular dynamics has been used extensively to study protein-ligand interactions~\cite{heifetz2020quantum,Takaya2021} (including HIV-1 binding to Darunavir~\cite{Chuntakaruk2024}). 

\subsection{Multiple Supercomputer Platform}

In our work, we use a combined approach of two supercomputers, one with standard CPUs, the second with GPU accelerators. This type of heterogeneous supercomputer can be seen as an extension to previous practices where a GPU partition may have been prepared for visualization purposes. An example of such an offering is the Wisteria/BDEC-0 supercomputer at the University of Tokyo~\cite{wisteria2024}. This machine offers two powerful partitions linked by a shared file system: one composed of A64FX CPUs (Odyssey) and the other of A100 GPUs (Aquarius). On this system, one may even simultaneously reserve a job on both partitions, with communication between partitions handled seamlessly through MPI~\cite{sumimoto2022system}. As a related paradigm, we highlight the Tsubame supercomputer at Institute of Science Tokyo, which allows users to reserve only the CPU portion of its nodes, such that their calculations can overlap with another user that is primarily using the GPUs.
In this workflow, we run the MM portion on Aquarius, and the QM portion and post processing on Odyssey, inline with the available implementations of the underlying methods. 

The CPU partition of the Wisteria/BDEC-0 supercomputer (Odyssey) has 7,680 nodes of A64FX processers, each with 48 cores. The total Odyssey partition has 25.9 PFLOPs of FP64 performance and 7.8 PB/s aggregate memory bandwidth. The GPU partition (Aquarius) has 45 nodes, each with two Intel Xeon Platinum 8360Y processors and 8 NVIDIA A100 GPUs. The total Aquarius partition has 7.2 PFLOPs of FP64 performance 578.2 TFLOP/s of aggregate memory bandwidth. For GENESIS MM calculations, we utilize one full node of Aquarius (8 A100 GPUs). For the BigDFT calculations, we run each simulation on 64 Odyssey nodes. Post processing is performed on a single node of Odyssey. Previous examples of coupling together the two partitions of Wisteria/BDEC-0 include use in multiphysics climate simulations where only certain models have been ported to GPU~\cite{multi_nakajima} and simulating quantum-classical coupling~\cite{hamamura2026integrating}.

\subsection{Combined Workflow}

The workflow is orchestrated by the remotemanager package~\cite{Dawson2024}, which modifies arbitrary python functions so that they are executed on a remote computer through the jobscript system. An example of how remotemanager is used on the Wisteria/BDEC-0 machine is shown in Listing~\ref{lst:remote}. A computer is defined based on a template of the jobscripts for the machine and the parameters the user wishes to modify at a per job granularity (in the example case, the number of OpenMP threads). The user next defines the function they wish to run remotely and creates a dataset based on this function and computer definition.  The arguments for each run are set and then the jobs are run asynchronously. In this simple example, we wait for the full dataset to complete before fetching the results; however, as soon as any individual job completes its results are available.

\begin{lstlisting}[caption={Remote job submission to the Wisteria/BDEC-0 machine using remotemanager. This listing shows an example of how to transform a simple function into a dataset that is automatically submitted remotely.},label={lst:remote}]
from remotemanager import Computer, Dataset
# project and user variables defined elsewhere

# define the computer based on a template, url, and submitter
template = f"""#PJM -L rscgrp=regular-o
#PJM -L node=#NODES#
#PJM --omp thread=48
#PJM -L elapse=00:10:00
#PJM -g {project}
module load python"""

comp = Computer(template=template,
                host="wisteria.cc.u-tokyo.ac.jp",
                submitter="pjsub",
                )

# function to execute remotely (execute the computational workload)
def function_to_execute(**kwargs):
    import BigDFT
    return BigDFT.run(**kwargs)  # pseudocode

# create a dataset of runs of this function
ds = Dataset(function_to_execute, url=comp, remote_dir=f"/work/{project}/{user}")
ds.append_run({"xc": "PBE0"}, nodes=16)
ds.append_run({"xc": "PBE"}, nodes=1)

# submit asychronously, wait, fetch result
ds.run() ; ds.wait() ; ds.fetch_results()
print(ds.results)
\end{lstlisting}

\begin{figure}
    \centering
    \includegraphics[width=0.5\textwidth]{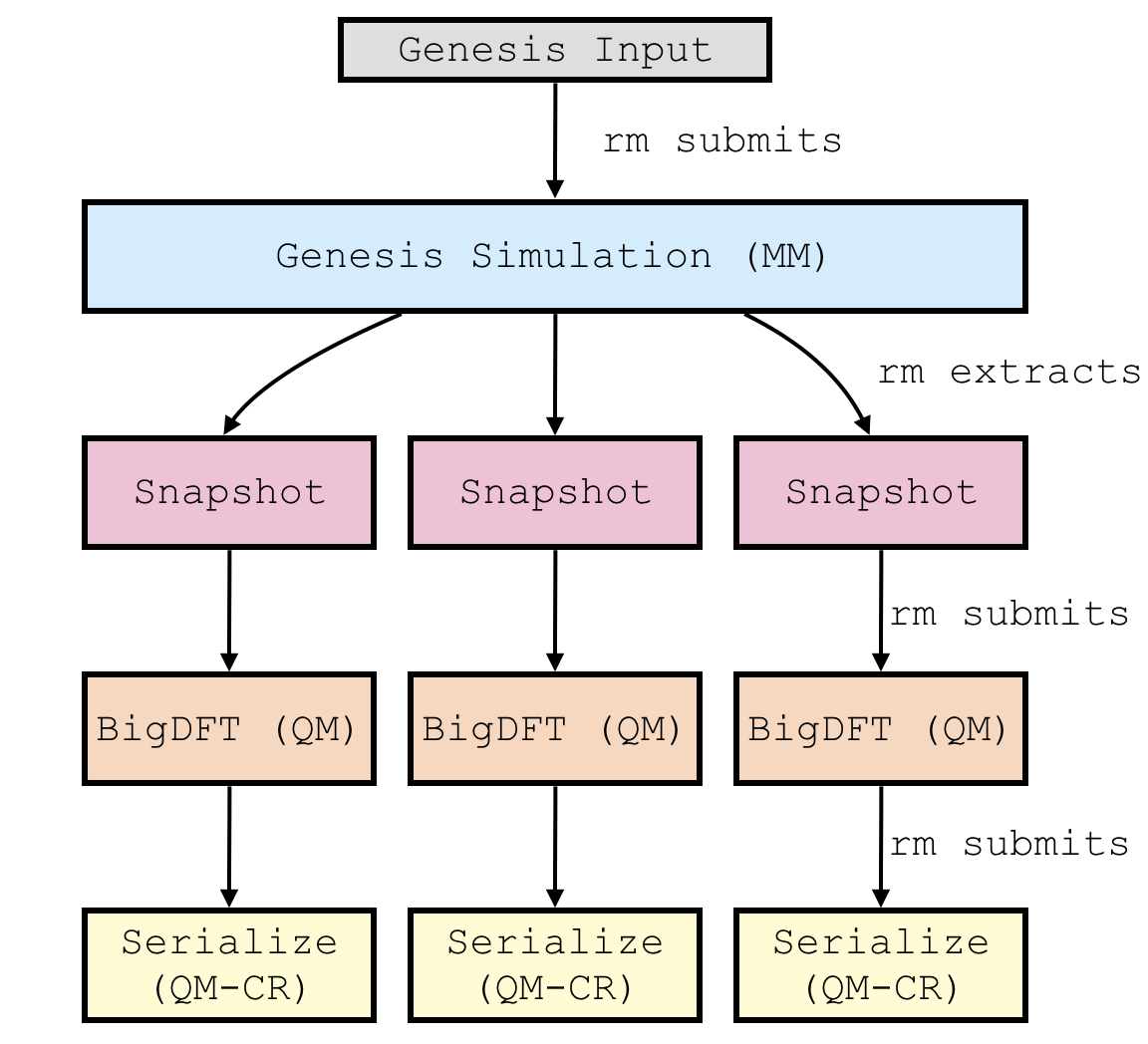}
    \caption{Overall coupling of the MD + QM + QM-CR workflow through remotemanager (rm). Four different datasets are generated with remotemanager for running GENESIS (MM), extracting snapshots, running BigDFT (QM), and serialization (QM-CR). Each dataset runs its job as soon as data becomes available.}
    \label{fig:coupling}
\end{figure}

In the case of our combined workflow, a manager process runs on the user's personal computer to orchestrate the coupling outlined in Figure~\ref{fig:coupling}. First, a GENESIS job is submitted that runs molecular dynamics, writing new trajectory files at a fixed frequency. A second subroutine is run periodically that checks for a new frame, using MDAnalysis~\cite{Michaud-Agrawal2011} to extract snapshots. If a new set of positions is available, a BigDFT calculation is generated and submitted. Once a BigDFT calculation is completed, a serialization step is submitted that uses QM-CR to post process the data into a form easy for analysis at a later stage. In principle the connection between BigDFT and serialization jobs could be managed through a scheduler's dependency management; however, the GENESIS job producing new data for calculation while it is ongoing is an example of where a manager process must be used. We generate separate workflows for different calculation stages based on the process in GENESIS:  optimization, equilibration, and production. For convenience, we break the production phase into two stages of 40~ns each.

\section{Results}


\subsection{Overlapping Workflow Effectiveness}

\begin{figure}
    \centering
    \includegraphics[width=1.0\textwidth]{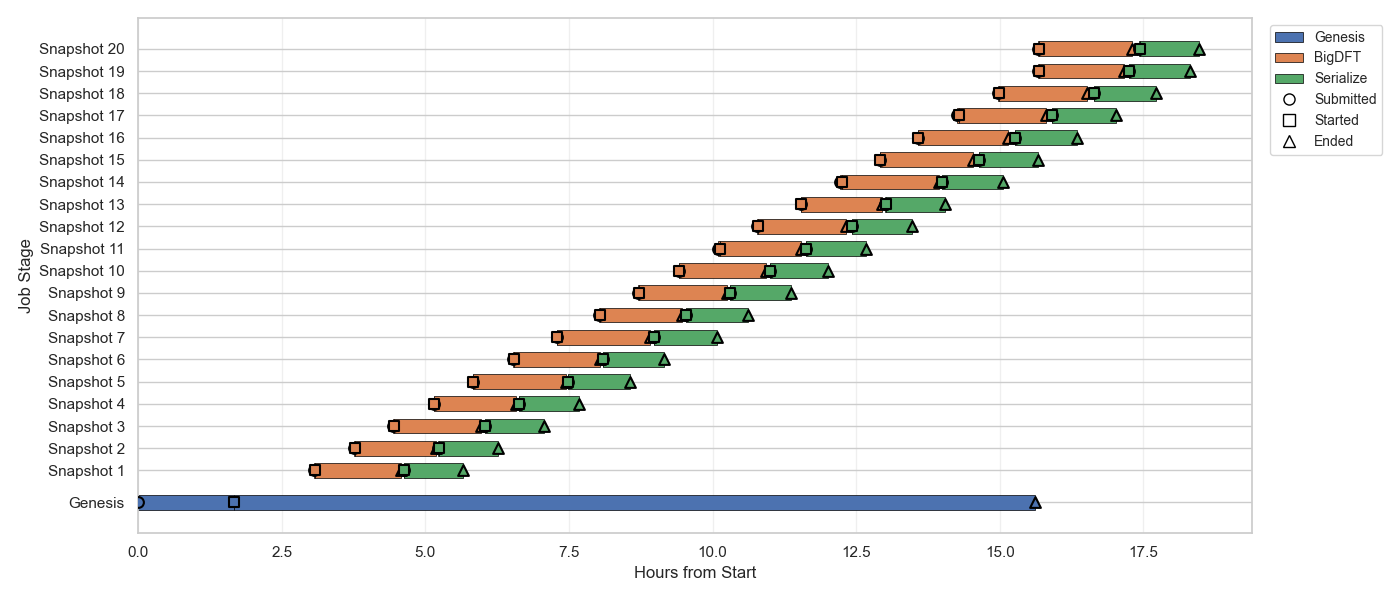}
    \caption{Timeline of calculations for the first production phase (40~ns) of the wild type variant. As the GENESIS calculation runs, snapshots are produced, which are transformed into BigDFT and serialization jobs.}
    \label{fig:prod1}
\end{figure}

We first investigate the effectiveness of our combined MD + QM + QM-CR approach at utilizing available computing resources. In Figure~\ref{fig:prod1} we show the timeline of job submission and execution for each part of the workflow. As we ran our calculation during a time of low system utilization, we see that the BigDFT and Serialization jobs began almost immediately after submission. The final two snapshots were launched at the same time due to the way an incomplete trajectory is processed by MDAnalysis, where the last written trajectory is unavailable until completion of the full run. Due to the specific problem size and compute partitions selected, we see an overlapping of up to three simultaneous BigDFT calculations, alongside multiple serialization calculations. We note that running all twenty BigDFT calculations simultaneously would require roughly 1/6th of the Odyssey partition; hence, eagerly submitting calculations as soon as they become available is advantageous for throughput under practical machine load where such a large reservation is unlikely to be available at any given moment.

\subsection{Base Structure of Darunavir: MD-coherent Fragmentation}

\begin{figure}
    \centering
    \includegraphics[width=1\textwidth]{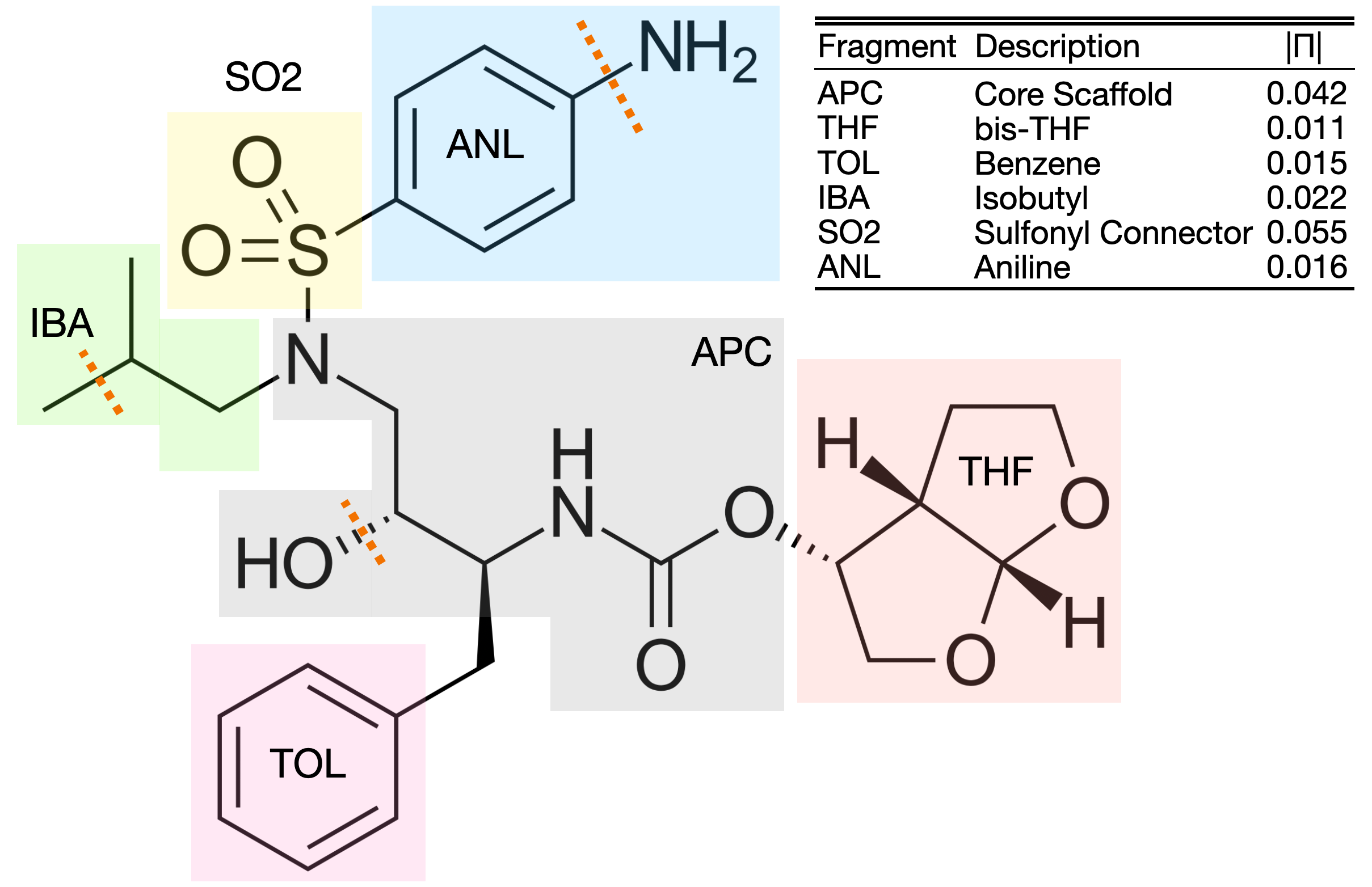} 
    \caption{Fragmentation of DRV using the QM-CR purity indicator. The top right table assigns an abbreviation to each fragment, describes the fragments in terms of chemical groups, and reports the purity indicator ($|\Pi|$) from a gas phase calculation. The Purity Indicator identifies the independent modular elements through which DRV works. Therefore, it allows us to identify which specific chemical groups may be affect by mutation-induced loss of binding. The orange dashed lines represent the additional fragmentation that would emerge when using a looser cutoff value of $0.1$.}
    \label{fig:fragmentation}
\end{figure}

Before analyzing the full complex, we focus only on DRV in order to decompose it into the right level of granularity for analysis. QM-CR defines a measure called the purity indicator ($|\Pi|$), which uses the density matrix to determine the degree of confidence in which a given set of atoms can be treated as an independent fragment. By combining this indicator with an optimization algorithm, we can construct from first-principles a reliable fragmentation of a given molecule (see our previous work~\cite{10.1021/acs.jctc.9b01152,Genovese2023}, as well as the Appendix, for more details). The protein itself is decomposed on the basis of its amino acid sequence --- which reliably have $|\Pi|$ values below $0.075$ --- giving us a means of treating the protein and ligand at the same level of detail.
This unified resolution ensures the protein is not analyzed as an approximated environment, but as a chain of well-identified interaction units. This resolution allows for pinpointing the exact residue-level coordinates where the drug's binding is affected, while contextually characterizing how the virus trades its own binding strength in view of evading inhibition.
In Figure~\ref{fig:fragmentation} we show the fragmentation that we use along with the naming scheme. This fragmentation was selected by first performing an automatic fragmentation with a loose cutoff point of $0.1$. This resulted in a similar fragmentation, but with the IBA, APC and ANL fragments split into two. As both had high $|\Pi|$ values and the fragmentation arbitrarily split one methane group in IBA as opposed to the other, we merged these fragments, resulting in $|\Pi|$ values below a cutoff of $0.075$.

\begin{figure}
    \centering
    \includegraphics[width=1\textwidth]{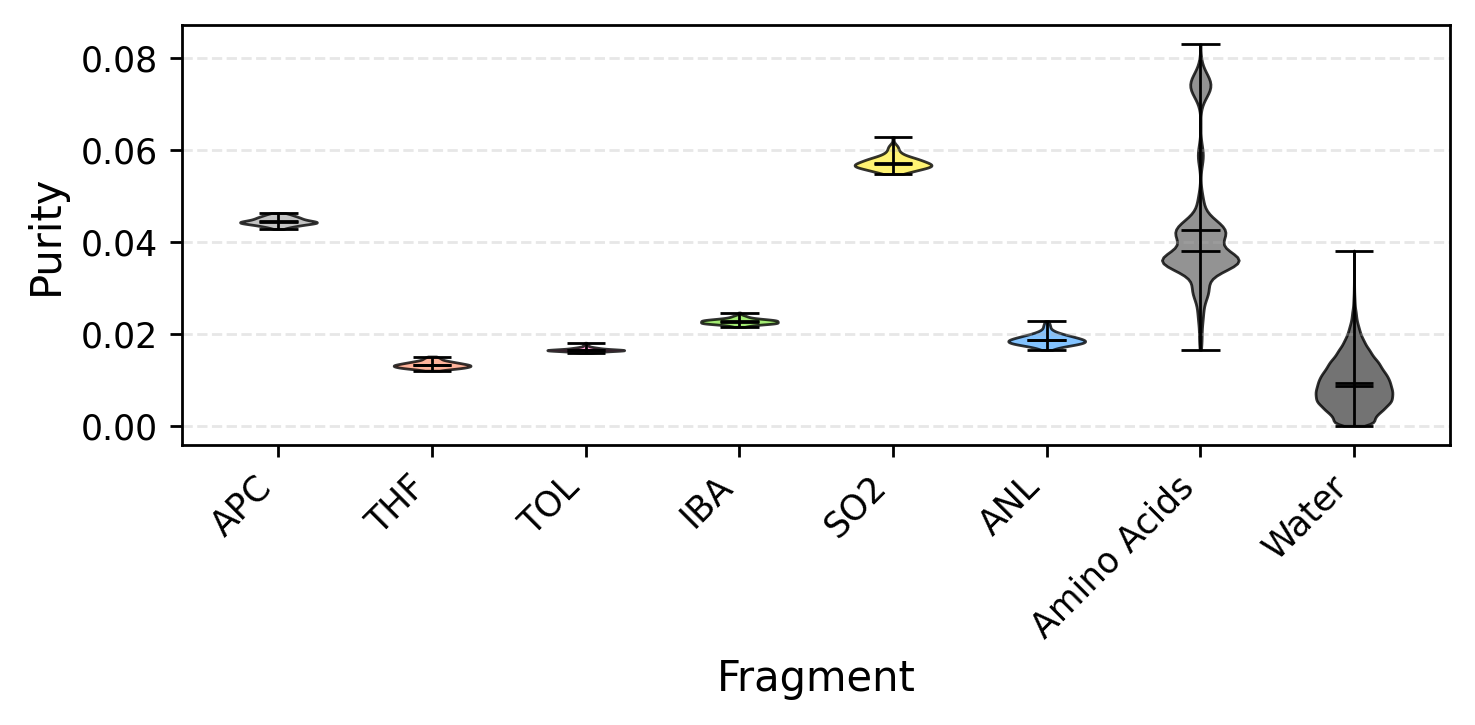} 
    \caption{Distribution of the purity values of different inhibitor fragments (DRV), protein amino acids, and water molecules over the course of the production runs.}
    \label{fig:pi_distribution}
\end{figure}

 In Fig.~\ref{fig:pi_distribution} we plot the variability of the $|\Pi|$ values of the DRV fragments and protein amino acids over the course of the molecular dynamics runs. In general, the $|\Pi|$ values increase compared to the gas phase results due to interactions with PR. The largest amount of variability is in the SO2 and ANL fragments (fragments that will be discussed in Sec.~\ref{sec:loss}). Such an approach provides a coherent, non-arbitrary method to define systems' moieties on an equal footing. As all fragments have a $|\Pi|$ value below the amino acid threshold for the entirety of the trajectories, we can interpret this fragmentation as a time-preserved coarse-grained view of the system.


\subsection{Mutation Induced Global Modification of Electronic Interactions}

To quantify the effect of mutations on interactions between DRV and PR, we first analyze the statistical change in interaction terms obtained from the fragment-based electronic decomposition. The QM-CR framework provides two measures of fragment interaction: the electrostatic interaction energy ($E_{\mathrm{el}}$) and the fragment bond order (FBO). The first term describes long range interactions induced by charge centers and the second the short range chemical interaction (see Appendix). We compute these measures for each frame of the molecular dynamics trajectories. The quantities reported below correspond to the statistical distributions of the differences between the mutant trajectories and the wild-type (WT) ensemble. For example, the change in the DRV-PR electrostatic interaction between the WT and mutant $M$ is:
\begin{equation}
\Delta E^{\mathrm{DRV-PR}}_{\mathrm{el}} = \sum_{g \in T_{\mathrm{WT}}} \sum_{f \in T_M} 
\left[E_{\mathrm{el}}^{\mathrm{DRV-PR}}(f) - E_{\mathrm{el}}^{\mathrm{DRV-PR}}(g) \right],
\end{equation}
where $f,g$ are snapshots from a trajectory $T$. Such quantity can also be interpreted as the statistical population of the interaction loss. A positive value of this quantity will therefore imply that the mutant $M$ has a less strong electrostatic binding to the DRV drug than the original WT PR. The same statistical distribution can be calculated for the FBO terms, characterizong the chamical entanglement of the DRV-PR interface. In that case, as the FBO is always a positive defined quantity, an interaction loss would be then associated to a negative value of $\Delta FBO^{\mathrm{DRV-PR}}$.

\begin{figure}[htbp]
\centering
\includegraphics[width=0.32\textwidth]{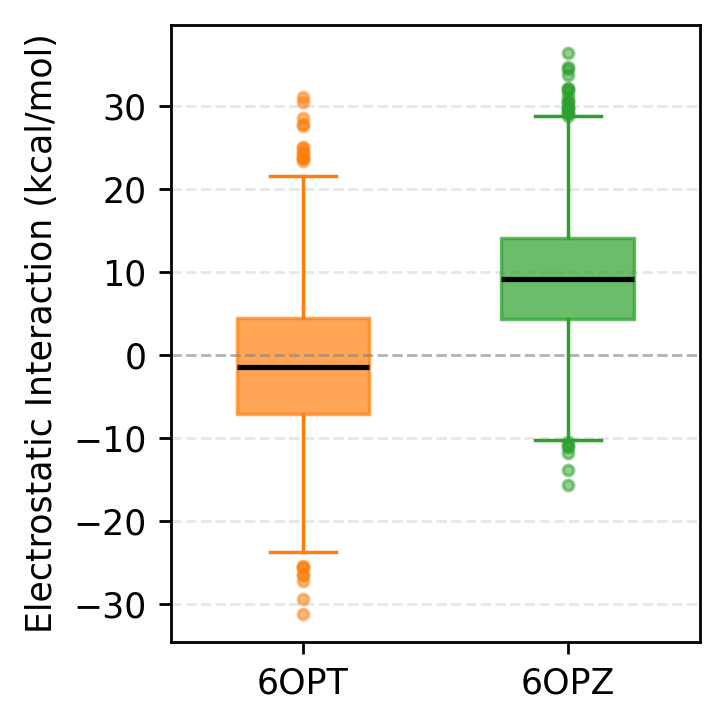}
\includegraphics[width=0.32\textwidth]{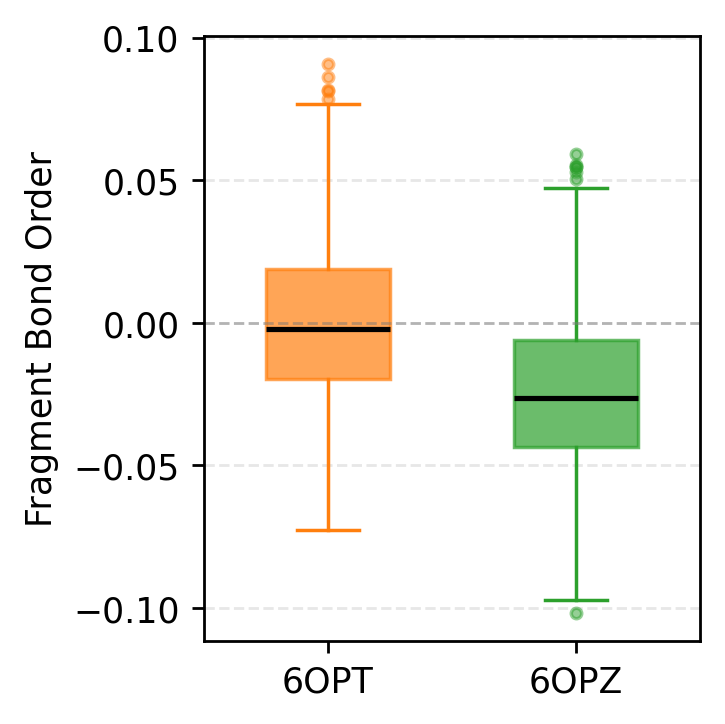}
\caption{Global modifications of interaction terms between DRV and PR upon mutation. Left: change in the electrostatic interaction energy. Right: change in the fragment bond order. Distributions are obtained from the difference of ensemble production MD configurations between the mutated strand and the WT system. A weaker interaction (loss of affinity) is indicated by a more positive value for the electrostatic interaction and a more negative value for the FBO.}
\label{fig:global_interactions}
\end{figure}

Figure~\ref{fig:global_interactions} summarizes the distributions of the total changes in $E_{\mathrm{el}}$ and FBO for the mutated systems relative to the WT reference. Both measures display broad distributions across the MD trajectories, reflecting the significant conformational variability of the DRV–PR complex. 
Despite these fluctuations, systematic trends emerge when comparing the WT and mutant ensembles. In particular, the 11 mutation 6OPZ system exhibits a shift of all interaction terms toward less stabilizing values, i.e. weakening overall DRV binding. The electrostatic term becomes more positive indicating a reduction in attractive or increase in repulsive interaction terms. 
The fragment bond order provides a complementary view of the changes of binding. As FBO measures the portion of the electronic density shared between fragments, its decrease in the 6OPZ snapshots indicates a reduction of electronic coupling between DRV and the surrounding residues.



The 6OPS (2 mutation) system exhibits smaller deviations from the WT distributions, suggesting that the electronic interaction network between DRV and the protease remains largely preserved in this case. In contrast, the more extensive mutation pattern of the 6OPZ system induces a measurable redistribution of both electrostatic and short-range electronic interactions, consistent with the substantial decrease in DRV affinity observed in experiment.


\subsection{Interaction Fingerprints}
\label{sec:finger}

\begin{figure}[htbp]
\centering
\includegraphics[width=1.\textwidth]{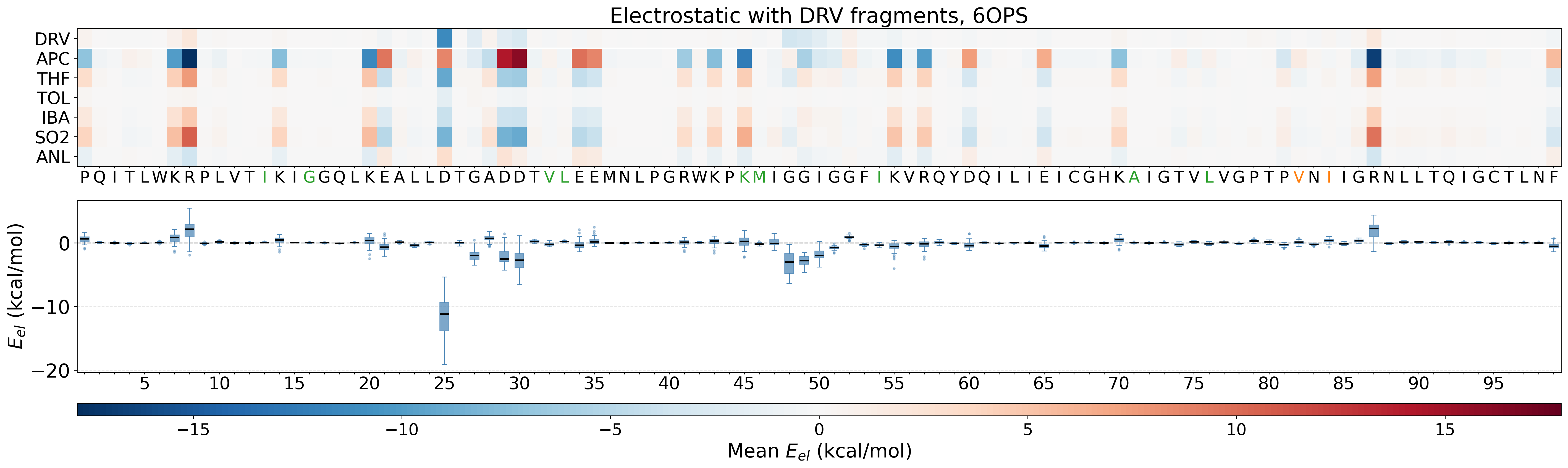}\\
\includegraphics[width=1.\textwidth]{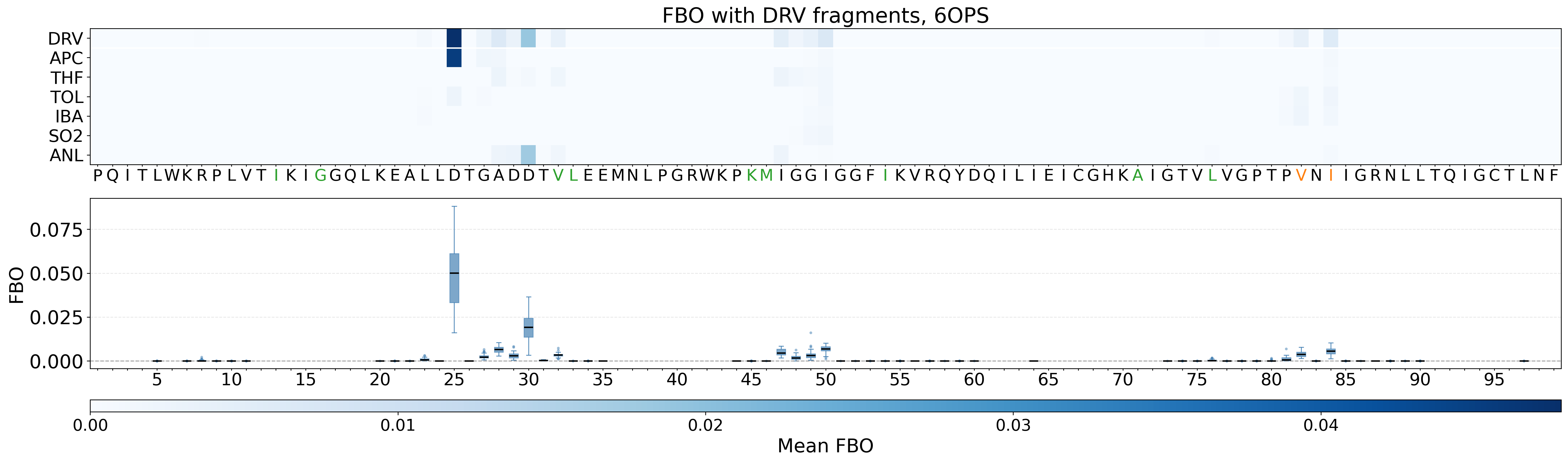}
\caption{Residue-resolved $E_{\mathrm{el}}$ energy (top) and FBO (bottom) between DRV and PR in the WT complex. The average interactions are resolved at a per DRV fragment level (heatmap); we also plot the distribution of interactions for the full DRV molecule. The distributions come from the ensemble of MD configurations and reveal the localization of salient regions along the WT protease sequence. Residues mutated in 6OPT and 6OPZ are written in orange and green, respectively.}
\label{fig:residue_contribution}
\end{figure}

To identify the regions of the protease that contribute the most to the interaction with DRV, we analyze the residue-resolved electrostatic interactions and fragment bond orders for the WT complex (Figure~\ref{fig:residue_contribution}). Similar charts for the other two variants are available in the Supporting Information. The resulting quantities provide a sequence-resolved description of the interaction network that stabilizes the inhibitor within the binding pocket. For this work, we merge interactions between the two chains of PR for visual simplicity. The distributions correspond to the ensemble of configurations sampled along the production molecular dynamics trajectory and therefore reflect the statistical variability of the electrostatic interactions within the binding pocket.


The electrostatic profile reveals that residues located in the vicinity of the catalytic dyad (around positions 24–30) and the flap hinge region (around residue 45) display prominent interactions with the inhibitor. These regions are known to play a central role in stabilization of the inhibitor within the active site. The FBO distributions exhibit a spatial pattern that closely follows the electrostatic map. The consistent panorama of interactions between the electrostatic and FBO picture reinforces that DRV works by filling the substrate envelope to unlock short range interactions~\cite{L2010}.

The strongest electrostatic interactions arise from the central fragments of the inhibitor, in particular the APC and SO2 groups, which interact directly with residues located in the catalytic region of the protease. These fragments are therefore responsible for a large fraction of the electrostatic stabilization of the complex. Other fragments, such as THF and ANL, display more moderate electrostatic contributions, while IBA and TOL exhibit relatively small electrostatic interactions. The APC fragment also displays the largest FBO contributions, indicating that it acts as a central electronic anchor for the inhibitor within the binding pocket. The THF and ANL fragments also contribute to the electronic coupling, although to a lesser extent, while fragments IBA SO2, and TOL exhibit relatively small bond orders.

\begin{figure}[htbp]
\centering
\includegraphics[width=\textwidth]{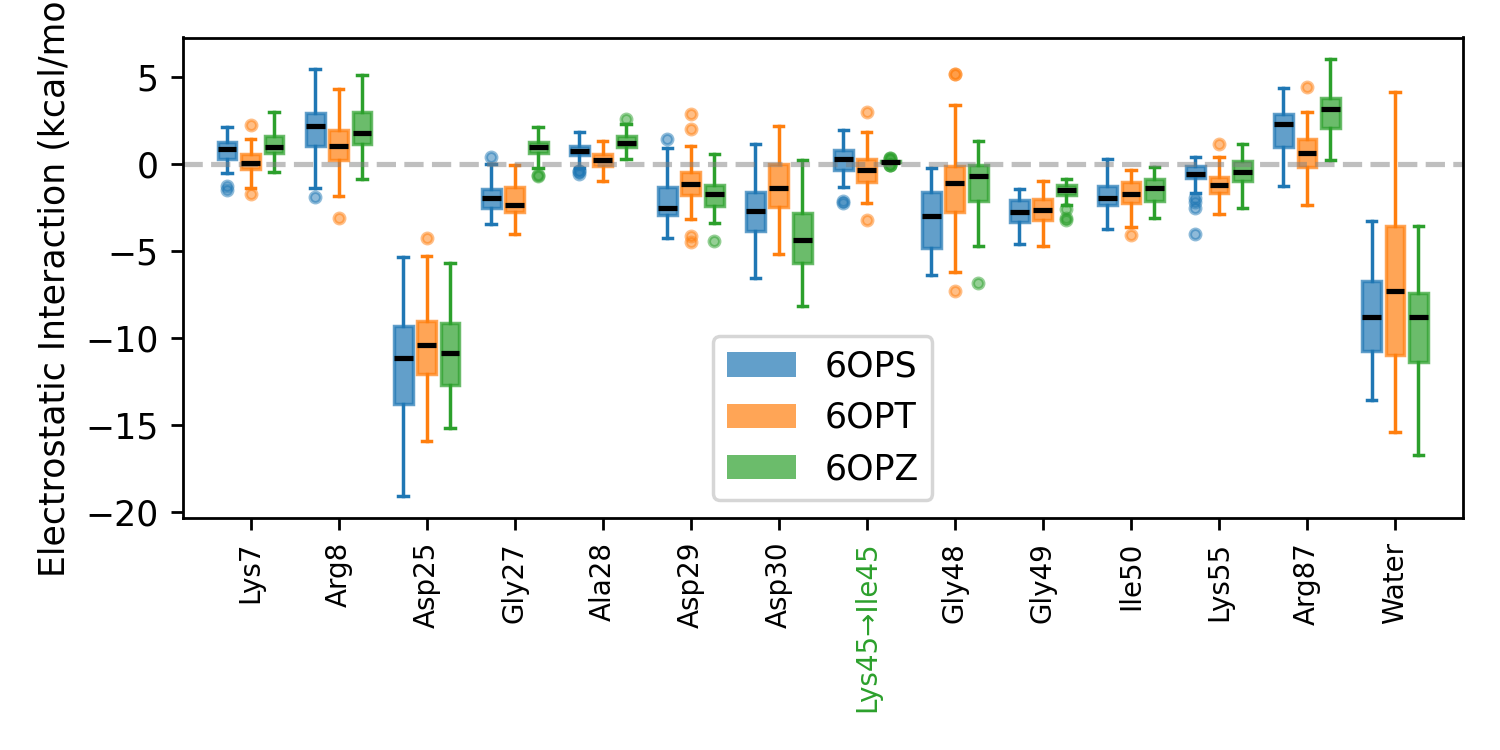}\\
\includegraphics[width=\textwidth]{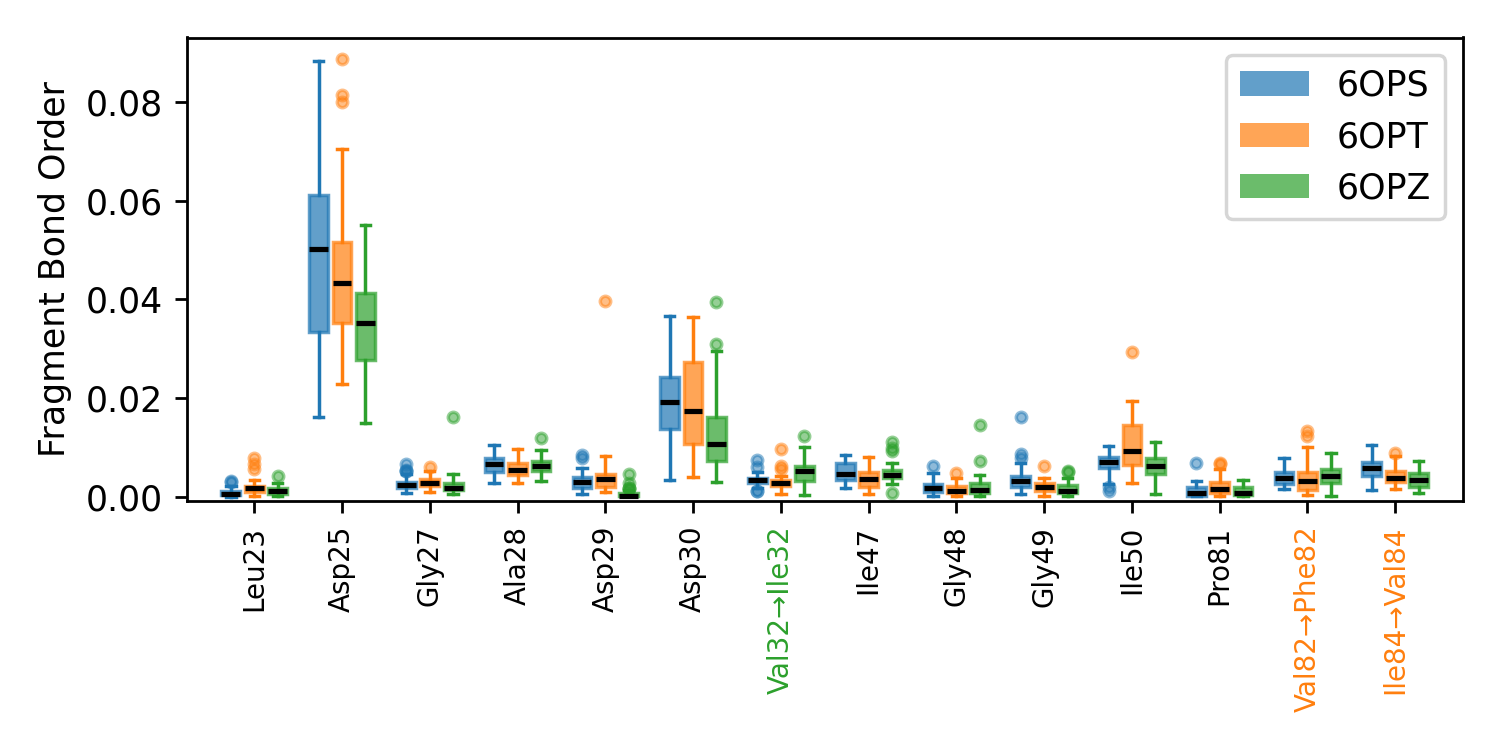}
\caption{Distribution of the electrostatic interaction energy (top) and FBO (bottom) at an amino acid level for the three mutants. We focus on the most significant interactions using an absolute cutoff of 1 kcal/moly and 0.001 for the electrostatic and FBO interactions, respectively. Residues mutated in 6OPT and 6OPZ are written in orange and green, respectively.}
\label{fig:PR_sidebyside}
\end{figure}

We plot a comparison of these descriptions between the variants in Fig.~\ref{fig:PR_sidebyside}. The main change in electrostatic interactions come from glycine residues either located near the active site (G27) or in the flaps region (G48 and G49). The prominence of glycine residues in the map shows how even backbone interactions, of substantial interest for inhibitor design~\cite{Ghosh2008}, can be impacted by a collection of mutations elsewhere.

For the 2 mutation variant (6OPT), we see a weakening of the FBO interaction at the mutated sites, as well as those in the area of the flaps I47, G48, and G49. The median FBO interactions of catalytic residue Asp25 is also decreased, but remains within the first and third quartile of the original interaction. For the flap interactions, there is a large increase in the FBO strength of residue I50. At the mutation sites V82 and I84, there is a reduction in FBO interaction only at I84. From this analysis, it would be difficult to predict that DRV has a weaker inhibitory effect on 6OPT; however, the results suggest that the modified binding pattern with the flap region may contribute to the change. For the 11 mutant variant, however, we see much greater FBO weakening in the area of the catalytic triad, particularly at D25, D29, and D30. The only compensating interaction strengthening over the wild type is at V32. This interaction shows a counterintuitive mutation mechanism: rather than weakening an existing binding motif, the interaction with V32 is strengthened to reduce binding elsewhere. 

\subsection{Inhibitor Level Changes for the 11 Mutation System}
\label{sec:loss}
We have seen that the 11 mutation system exhibits a significant depletion of binding compared to the WT. To understand how these changes emerge from distal mutations, we analyze how the variations in electrostatic interaction and FBO are distributed among the different fragments of DRV. 

\begin{figure}[htbp]
\centering
\includegraphics[width=\textwidth]{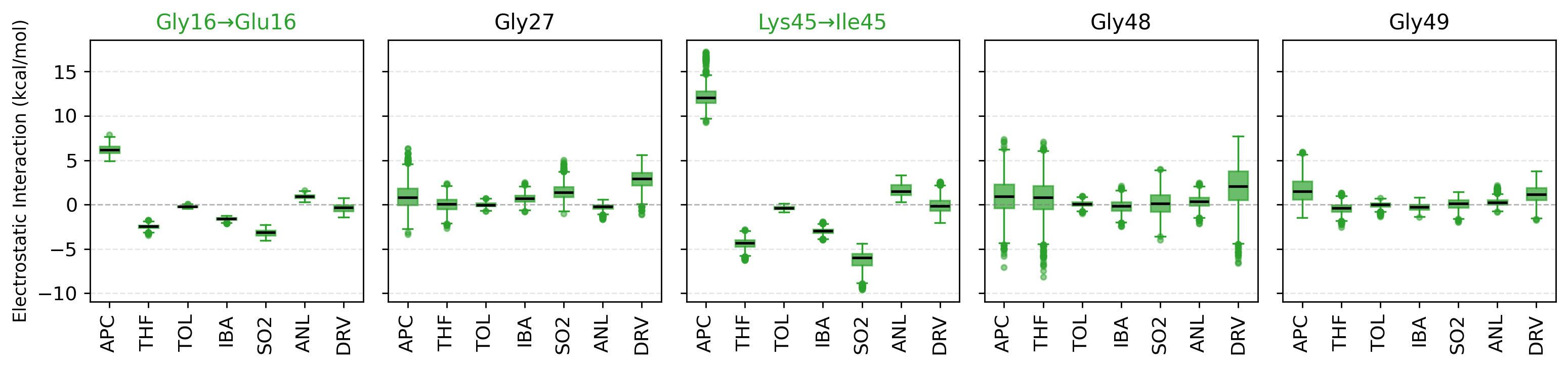}\\
\includegraphics[width=0.6\textwidth]{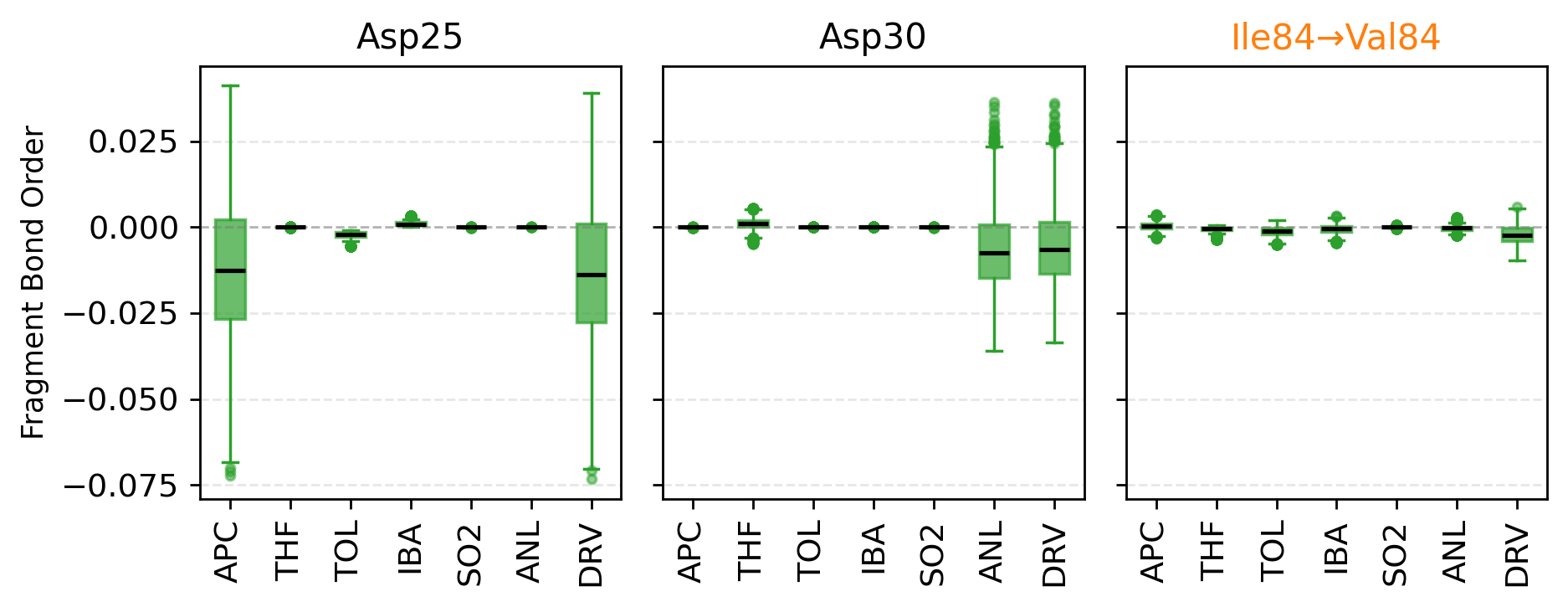}
\caption{Fragment decomposition of the mutation-induced loss in electrostatic interaction energy (top) and FBO delta (bottom) for selected protease residues of the 11 mutation system. Each panel shows the contribution of individual DRV fragments to the interaction difference between the 6OPZ and 6OPS systems. Residues mutated in 6OPT and 6OPZ are labeled in orange and green, respectively.}
\label{fig:fragmented_delta}
\end{figure}

Figure~\ref{fig:fragmented_delta} reports the redistribution of fragment-resolved interactions, focusing on a set of residues informed by the interaction pattern that emerged in Sec.~\ref{sec:finger}. Plots of all significantly modified interactions (including interactions with water molecules) are available in Supplementary Information. The redistribution of electrostatic interactions across residues is largely mediated by the central fragments of the inhibitor. In particular, variations in the electrostatic interaction associated with residues located in the catalytic region and the flap hinge primarily are driven by the SO2 and APC fragments. These fragments therefore act as the main electrostatic interaction centers of the ligand, transmitting the mutation-induced perturbations across the binding pocket.

The FBO decomposition indicates that the mutation-induced reduction of electronic coupling is primarily localized within the central scaffold of DRV. Specifically, the decrease in fragment bond order for residues in the catalytic region is most pronounced at the APC fragment. This result is consistent with the fragment-level analysis in the previous section, which identified the APC fragment as the primary site of electronic coupling for the inhibitor within the binding pocket.
 Thus, the combination of mutations work together to indirectly weaken the primary interaction point of DRV. There is also a significant decrease in interactions with ANL. Modification to the ANL group (for which there is greater freedom than modifying the core scaffold) thus represents a potential route to overcome resistance in the 11 mutation variant.



\subsection{Interaction Graphs}

\begin{figure}[htbp]
  \centering
  \begin{subfigure}[b]{0.65\textwidth}
      \centering
      \includegraphics[width=\textwidth]{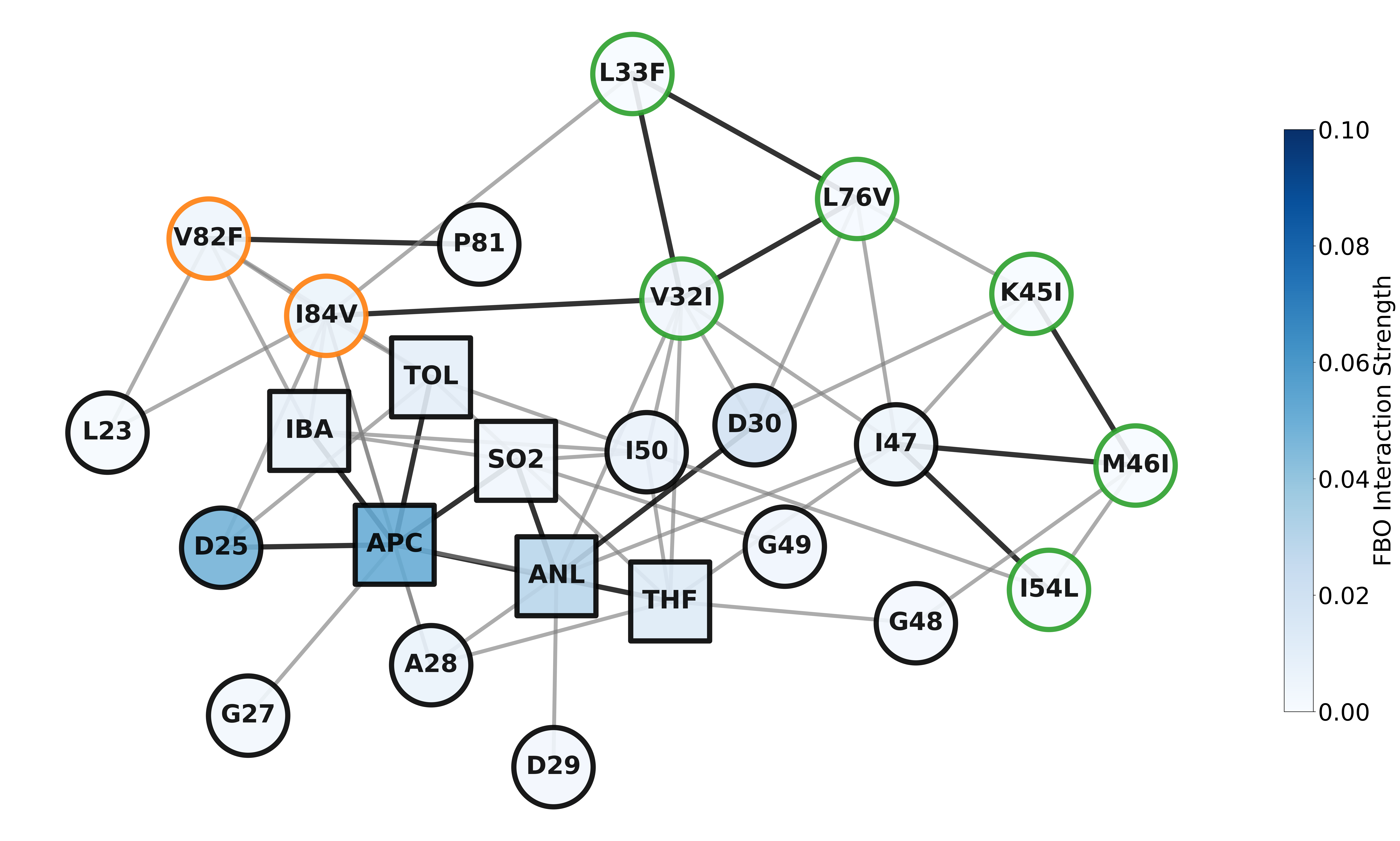}
      \caption{6OPS (Wild Type)}
      \label{fig:6ops}
  \end{subfigure}
  \vfill
  \begin{subfigure}[b]{0.65\textwidth}
      \centering
      \includegraphics[width=\textwidth]{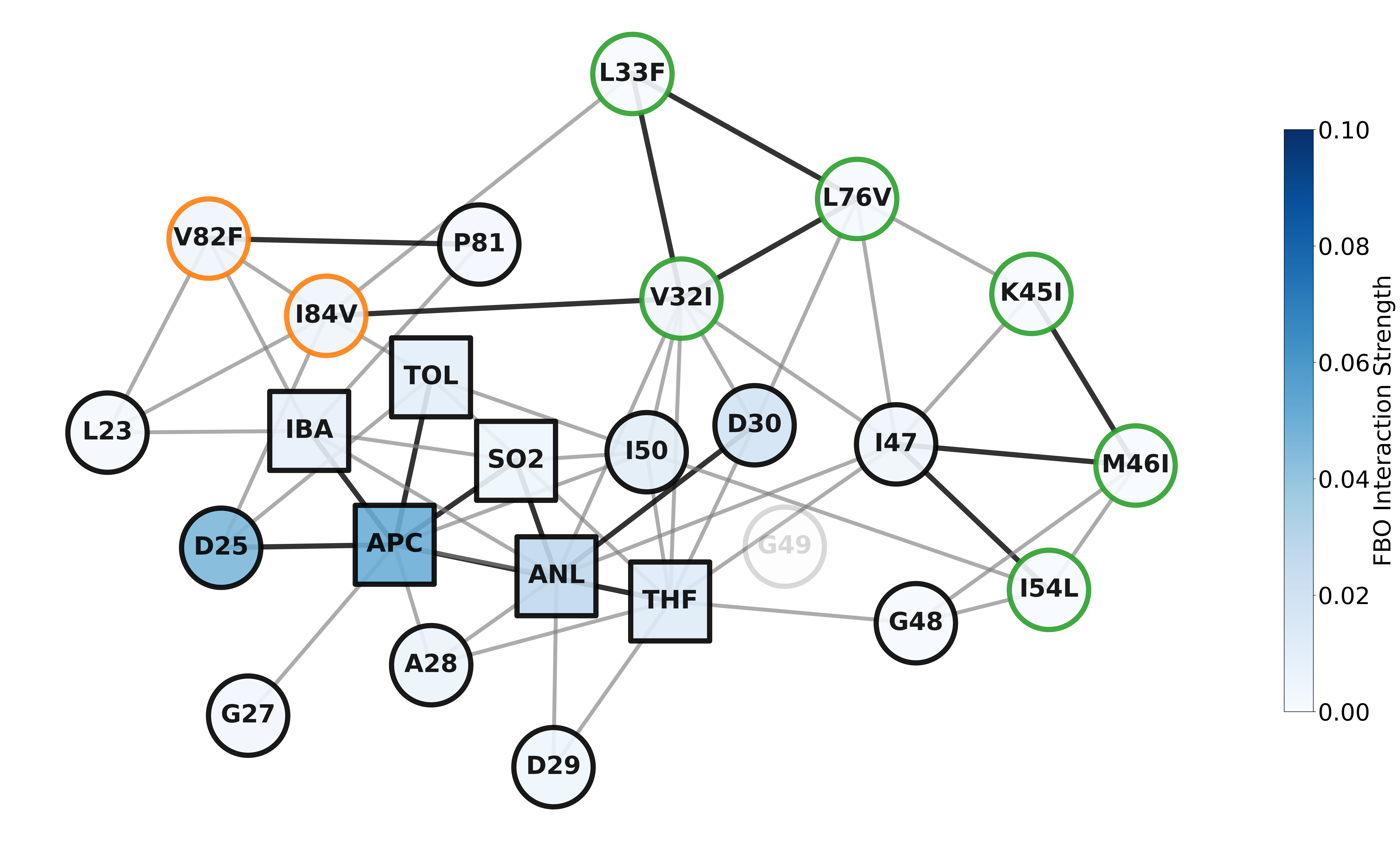}
      \caption{6OPT (2 Mutations)}
      \label{fig:6opt}
  \end{subfigure}
  \vfill
  \begin{subfigure}[b]{0.65\textwidth}
      \centering
      \includegraphics[width=\textwidth]{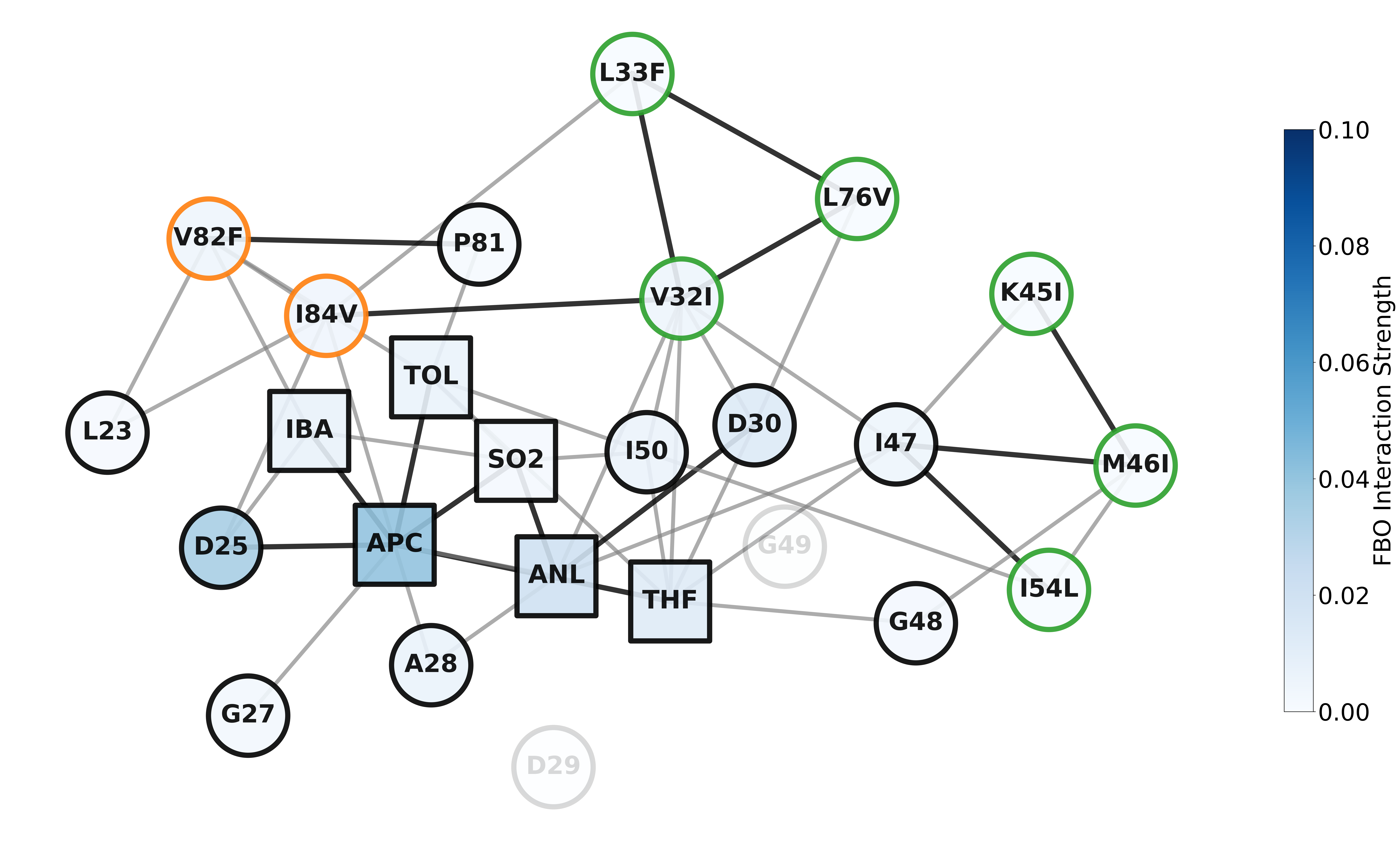}
      \caption{6OPZ (11 Mutations)}
      \label{fig:6opz}
  \end{subfigure}

  \caption{FBO interaction networks across the three variants. Node colors indicate FBO interaction strength with DRV. DRV fragments shown as squares, protein residues as circles. Colored outlines indicate mutations: orange for residues mutated in 6OPT, green for residues mutated only in 6OPZ. Edges are drawn as thick lines or thin lines if the FBO is greater than 0.01 or 0.001, respectively. Faded nodes are unconnected in a given variant.}
  \label{fig:network}
\end{figure}

From the FBO interaction network, we can draw a graph-like view of the assembly (Fig.~\ref{fig:network}). Fragments that share a non-negligible FBO are connected by a graph edge and color-coded with the relative importance that the fragment has with respect to binding. To define the graph, we use the average FBO from the production phase runs. We represent all nodes that are directly connected to any DRV fragments. We further include in the graph all mutated residues that are within two hops of a DRV fragment.

From this interaction point of view, we note the centrality of mutated residues V82 and I84, both of which are connected to two fragments of DRV. We also note how distal mutations L33, I54, and L76 form a star topology around the proximal mutation V32. K45 interacts strongly with D30 in the wild-type and 2 mutation variant, yet after mutation this interaction disappears. We further note that I13 and M46 are connected in the graph by one more hop through interaction with I54 and L33, respectively (thus they are not shown in the graph). Hence, while this set of mutations may be ``distal'' in terms of spatial proximity to the active site, in the interaction network their relevance is already clear from the wild type structure.

\subsection{Stability of the Interaction Network}
\label{sec:equil_check}

\begin{figure}[htbp]
\centering
\includegraphics[width=\textwidth]{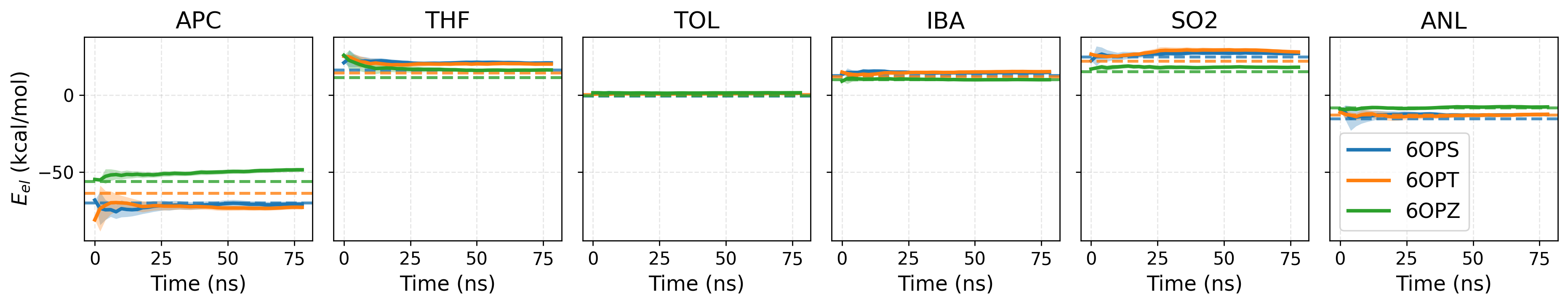}\\
\includegraphics[width=\textwidth]{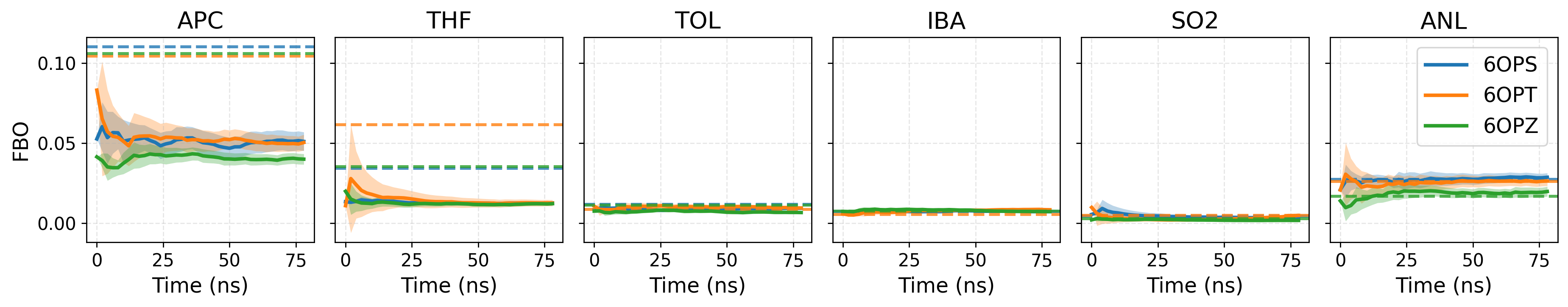}
\caption{Running cumulative mean of interaction quantities (electrostatic interaction and fragment bond order) for different fragments of DRV. Dashed horizontal lines indicate the values computed from the crystal structures after the initial optimization phase. The shaded areas represent the 95\% confidence interval.}
\label{fig:mean}
\end{figure}

The analyses above are derived from ensembles of configurations sampled along the MD trajectories. It is therefore important to assess whether the interaction patterns identified in the previous sections are stable over time or whether they arise from transient structural fluctuations. To address this point, we analyze the temporal evolution of the fragment-resolved interaction quantities along the trajectories (Fig.~\ref{fig:mean}). The time-resolved interaction profiles show that the relative contributions of the different ligand fragments remain remarkably stable throughout the simulations. While thermal fluctuations produce moderate variations in the instantaneous interaction values, the hierarchy of fragment contributions remains largely unchanged.

An additional insight is provided by the comparison between the crystallographic pose and the ensemble of MD configurations. For several fragments, the interaction values associated with the crystallographic structure lie close to the median of the MD distributions, indicating that the experimental pose provides a reasonable representation of the dominant interaction pattern. The main exceptions are the weakening in FBO of interactions with APC and THF. These observations indicate that while the crystallographic structure provides a useful structural reference, the interaction network between DRV and the protease should be interpreted as a dynamic ensemble rather than a single static configuration. The MD trajectories therefore provide essential information about the range of interaction strengths that contribute to the stabilization of the inhibitor within the binding pocket.

\section{Discussion}

From a computational standpoint, this work demonstrates the feasibility and advantage of a hybrid MD + QM workflow deployed across heterogeneous CPU/GPU architectures. The workflow leverages the high-throughput nature of GPU-based MD to generate conformational ensembles and immediately subjects selected frames to in-operando QM calculations on CPU nodes using a linear-scaling DFT implementation. This on-the-fly processing allows for continuous analysis without the need to halt production runs or transfer large datasets between clusters. This data-flow-oriented workflow may go beyond the simple processing of regular MD steps to encompass advanced sampling techniques~\cite{Yang2019} or calculations generated by feature tracking algorithms. We may also imagine coupling different kinds of QM techniques, such as excited state calculation of snapshots to reveal the statistical dispersion of electronic properties over a given MD trajectory. These techniques may be enabled by the capability of remotemanager to combine different computational resources.


By coupling extensive classical molecular dynamics with quantum mechanical analysis of trajectory frames, we observe a systematic weakening of the interaction between the DRV molecule and the PR in the 11 mutation variant. Weakening primarily comes from interaction with the backbone of DRV (APC), which represents a tradeoff for HIV-1 between inhibition by DRV and catalytic efficiency. We observe as well a short range weakening between ANL and D30, which is an interaction that may be targeted when designing new drugs using the DRV scaffold. The electrostatic interactions also reveal a weakening between APC and mutation point G16E and indirect interactions to G27 in the catalytic region and G48 and G49 in the flaps.


In the same APC-D25 interaction region, we observe a decrease in the interaction for the 2 mutation variant; however, the overall change is small to the point where we would be unlikely to predict this result in-silico using our current computational setup. Our data suggests that the decrease in inhibitory power may be connected to the change in interaction patterns at the flaps. Examining the full network of interactions defined by the FBO revealed that mutations L33, K45, I54, and L76 are not so distal after all, being only one hop away from key interacting residues (with I13 and M46 being only one more hop away). Our analysis did not reveal any specific interactions with A71 --- a residue that is completely unconnected to the other relevant amino acids in the graph view --- which is consistent with the small change in binding observed by introducing this mutation in the study of Henes et al.

Beyond this case study, the proposed framework opens the way to predictive modeling of drug resistance and protein-ligand interactions at scale. By reducing the overhead associated with post hoc QM analysis and enabling early detection of meaningful structural and electronic changes, our workflow supports rapid prototyping and evaluation of novel mutations or inhibitors. The flexibility of the approach also allows for generalization to other biological systems where distal mutations or allosteric effects complicate structure-activity relationships. 
As the development of portable and scalable quantum methods continues, we anticipate that workflows such as the one presented here will become increasingly central to computational drug discovery. This shift from a structural-driven assessment to an explicit mapping of electronic coupling redefines the ``druggable'' landscape of the protease. By identifying the specific chemical groups where interaction loss is concentrated, this framework provides a prescriptive model for lead optimization.

\section*{Acknowledgements}

We thank Kengo Nakajima for discussions and support using the Wisteria supercomputer. Computer time was provided by the Joint Usage/Research Center for Interdisciplinary Large-scale Information Infrastructures (jh210022). This project received funding from the European High Performance Computing Joint Undertaking (EuroHPC JU) under the European Union’s Horizon Europe framework program for research and innovation and Grant Agreement No. 101136269 (HANAMI Project). We also acknowledge the RIKEN-CEA collaborative framework, and the French Computer time allocation DARI-A0190312049.

\section*{Supporting information}

The entire computational setup, including all input parameters, is available on Gitlab: \\(https://gitlab.com/wddawson/genesis-bigdft-coupling).

The supporting information document includes: additional interaction heatmaps of the 6OPT and 6OPZ variants; plots of pairwise FBO interactions between DRV fragments and PR residues or water molecules; and electrostatic interaction pairs that show statistically significant modifications under mutation.

\appendix

\section*{Appendix: Fragment Level Interaction Descriptors}
\label{sec:qmcr_descriptors}

The QM-CR methodology exploits the density matrix representation of DFT calculations to analyze large systems in terms of interacting fragments. In the BigDFT implementation, the electronic structure is represented in a localized basis of \emph{support functions} that are optimized in situ to adapt to the chemical environment of each localization region. Within this representation, the Kohn--Sham density matrix $\hat F$ and the Hamiltonian $\hat H$ are expressed in the basis of support functions. The locality of this basis allows one to perform DFT calculations with computational scaling that grows linearly with the system size while preserving the full quantum description of the electronic structure. The QM-CR framework introduces fragment projection operators $\hat W_F$ that project observables onto the degrees of freedom associated with groups of atoms (fragments). Using these operators, fragment-resolved quantities can be obtained directly from the density matrix and Hamiltonian of the full system.

\paragraph{Fragment-projected density matrix}
Given a fragment $F$, the fragment-projected density matrix is defined as
\begin{equation}
\hat F_F = \hat W_F \hat F ,
\end{equation}
where $\hat F$ is the density matrix of the full system. The trace of this operator corresponds to the electronic population of the fragment,
\begin{equation}
n_F = \mathrm{tr}(\hat F_F).
\end{equation}
These projected quantities allow the definition of fragment descriptors that characterize both the electronic integrity of fragments and their mutual interactions.

\paragraph{Purity Indicator}
The \emph{Purity Indicator} (PI) quantifies the degree to which a chosen group of atoms behaves as an independent electronic subsystem. It is a measure of the deviation from idempotency of the fragment-projected density matrix:
\begin{equation}
\Pi_F =
\frac{1}{q^{(0)}_F}
\mathrm{tr}\left(
\hat F_F^2 - \hat F_F
\right),
\end{equation}
where $q_F^{(0)}$ is the charge of the bare fragment intended as the sum of the electrons carried by each of the atoms of the group.

The purity indicator can be interpreted as a measure of the \emph{fragment valence}, normalized by the electronic population of the fragment (making this quantity intensive). It is easy to see that such quantity is always negative, as the density matrix has eigenvalues less than or equal to one. Small values of $|\Pi_F|$ indicate that the fragment behaves approximately as an independent electronic subsystem and therefore that the chosen fragmentation provides a meaningful description of the electronic structure.
In practical applications, the purity indicator provides a quantitative criterion to assess the reliability of fragment observables such as fragment charges, dipole moments, and fragment interaction energies.

\paragraph{Fragment Bond Order}
The \emph{Fragment Bond Order} (FBO) quantifies the portion of the density matrix shared between two fragments $F$ and $G$. It is defined as
\begin{equation}
B_{FG} =
\mathrm{tr}
\left(
\hat F_F \hat F_G
\right).
\end{equation}
This quantity represents the electronic coupling between fragments and generalizes the Mayer bond order to the level of fragments. The fragment bond order therefore provides a direct measure of electronic delocalization between subsystems.


\paragraph{Algebraic relations between $\Pi$ and FBO}
Using the completeness relation for a full partition of the system,
\begin{equation}
\sum_G \hat W_G=\hat I,
\end{equation}
one has
\begin{equation}
\sum_G \hat F_G=\hat F.
\end{equation}
It follows that
\begin{equation}
\sum_G B_{FG}
=
\sum_G \mathrm{tr}(\hat F_F\hat F_G)
=
\mathrm{tr}\left(\hat F_F\hat F\right)
=
\mathrm{tr}(\hat F_F).
\end{equation}
Therefore,
\begin{equation}
\mathrm{tr}(\hat F_F)=B_{FF}+\sum_{G\neq F}B_{FG}.
\end{equation}
Using the definition of the purity indicator one obtains
\begin{equation}
q_F^{(0)}\Pi_F=-\sum_{G\neq F}B_{FG}.
\end{equation}
Hence the purity indicator can be interpreted as the negative sum of the bond orders connecting fragment $F$ to the rest of the system, normalized by the bare number of electrons of the fragment,
\begin{equation}
\Pi_F=-\frac{1}{q_F^{(0)}}\sum_{G\neq F}B_{FG}.
\end{equation}
For two fragments $F$ and $G$ merged into a larger fragment $H=F\cup G$, one has
\begin{equation}
\hat F_H=\hat F_F+\hat F_G,
\end{equation}
which leads to
\begin{equation}
q_H^{(0)}\Pi_H
=
q_F^{(0)}\Pi_F
+
q_G^{(0)}\Pi_G
+
2B_{FG},
\end{equation}
with $q_H^{(0)}=q_F^{(0)}+q_G^{(0)}$.
This relation shows explicitly how the bond order between two fragments contributes to the purity of the merged subsystem.
The fragment bond order therefore quantifies the electronic coupling that must be absorbed when two fragments are merged into a larger subsystem.

\subsection*{Evaluation of the Fragment Population Variance for an Independent-Particle State}
We derive here the expression of the variance of the fragment population in the case of an independent-particle state.
Let $\hat W_F$ be the one-particle projector onto fragment $F$, satisfying
\begin{equation}
\hat W_F^2=\hat W_F.
\end{equation}
The corresponding fragment population operator for an $N$-electron system is
\begin{equation}
\hat N_F=\sum_{i=1}^N \hat W_F(i),
\end{equation}
where $\hat W_F(i)$ acts on electron $i$.
Squaring this operator gives
\begin{equation}
\hat N_F^2
=
\left(\sum_{i=1}^N \hat W_F(i)\right)^2
=
\sum_{i=1}^N \hat W_F(i)^2
+
\sum_{i\neq j}\hat W_F(i)\hat W_F(j).
\end{equation}
Using the projector property, one obtains
\begin{equation}
\hat N_F^2=\hat N_F+\sum_{i\neq j}\hat W_F(i)\hat W_F(j).
\end{equation}
The second term in this equation is a two-body operator associated to the finding of two \emph{different} electrons in the fragment F.

The expectation value of the fragment population is
\begin{equation}
\langle \hat N_F\rangle = \mathrm{Tr}(\hat F \hat W_F).
\end{equation}
For an independent particle density matrix, the expectation value of a product of one-body operators acting on two different particles can be written as
\begin{equation}
\sum_{i\neq j}\langle \hat A(i)\hat B(j)\rangle
=
\mathrm{Tr}(\hat F \hat A)\,\mathrm{Tr}(\hat F \hat B)
-
\mathrm{Tr}(\hat F \hat A \hat F \hat B).
\end{equation}
Applying this identity with $\hat A=\hat B=\hat W_F$ yields
\begin{equation}
\sum_{i\neq j}\langle \hat W_F(i)\hat W_F(j)\rangle
=
\mathrm{Tr}(\hat F \hat W_F)^2
-
\mathrm{Tr}(\hat F \hat W_F \hat F \hat W_F).
\end{equation}
Therefore, the second moment of the fragment population is
\begin{equation}
\langle \hat N_F^2\rangle
=
\mathrm{Tr}(\hat F \hat W_F)
+
\mathrm{Tr}(\hat F \hat W_F)^2
-
\mathrm{Tr}(\hat F \hat W_F \hat F \hat W_F).
\end{equation}
Subtracting the square of the mean population,
\begin{equation}
\langle \hat N_F\rangle^2=\mathrm{Tr}(\hat F \hat W_F)^2,
\end{equation}
one obtains the variance, which is positive definite by definition:
\begin{equation}
\mathrm{Var}(N_F)
=
\langle \hat N_F^2\rangle-\langle \hat N_F\rangle^2
=
\mathrm{Tr}(\hat F \hat W_F)
-
\mathrm{Tr}(\hat F \hat W_F \hat F \hat W_F).
\end{equation}
This expression can be rewritten as
\begin{equation}
\mathrm{Var}(N_F)
=
\mathrm{Tr}(\hat F_F)-\mathrm{Tr}(\hat F_F^2).
\end{equation}
Recalling the definition of the purity indicator
\begin{equation}
\Pi_F=\frac{1}{q_F^{(0)}}\mathrm{Tr}(\hat F_F^2-\hat F_F),
\end{equation}
we obtain the relation
\begin{equation}
\mathrm{Var}(N_F)=-q_F^{(0)}\Pi_F.
\end{equation}
The purity indicator can then be seen as the negative of the charge fluctuation normalized by the bare number of electrons of the fragment.
This is a interesting result: if the fragment is perfectly pure, then the fragment population does not fluctuate. 
If the fragment is electronically entangled with the rest of the system, $\Pi_F < 0$ as the population variance is positive by definition.
So the PI is essentially the negative normalized population variance.
An alternative formula involving the FBO is
\begin{equation}
\mathrm{Var}(N_F)=\sum_{G\neq F} B_{FG}.
\end{equation}
Thus, the variance of the fragment electronic population is exactly equal to the total bond order connecting the fragment to the rest of the system.

\paragraph{Electrostatic interaction}
Fragments defined according to the purity criterion possess well-defined charge distributions that can be characterized by multipole moments derived from the fragment density.
These multipolar expansions allow one to evaluate the electrostatic interaction between fragments, which corresponds to the Coulomb interaction between the charge distributions associated with each fragment. Unlike other interaction terms, the electrostatic interaction can be either attractive or repulsive depending on the relative orientation and sign of the fragment charges.
At sufficiently large separations, and under the approximation of \emph{rigid fragment densities}, where the density of the system is approximated by the sum of the fragment-projected densities $\hat F_F$ and $\hat F_G$, the electrostatic term provides the dominant contribution to the fragment interaction energy.

For two fragments $F$ and $G$, the electrostatic contribution originates from the ionic–ionic and electronic–electronic Coulomb interactions contained in the DFT energy functional. The corresponding contribution can be written as
\begin{equation}
E_{\mathrm{elst}}^{FG}
=
\frac{1}{2}
\int d\mathbf r d\mathbf r'
\frac{
-\rho_{el}^F(\mathbf r)\rho_{el}^G(\mathbf r')
+
\rho_{ion}^F(\mathbf r)\rho_{ion}^G(\mathbf r')
}{|\mathbf r-\mathbf r'|},
\end{equation}
where $\rho_{el}^F$ and $\rho_{ion}^F$ denote respectively the electronic and ionic charge densities associated with fragment $F$.

\paragraph{Fragment density decomposition}

The electronic density of the system can be decomposed into fragment contributions,
\begin{equation}
\rho(\mathbf r) = \sum_{\mathcal F} \rho_{\mathcal F}(\mathbf r),
\end{equation}
where the fragment density itself can be expressed as a superposition of atomic densities,
\begin{equation}
\rho_{\mathcal F}(\mathbf r)
=
\sum_{a \in \mathcal F}
\rho_a(\mathbf r - \mathbf R_a).
\end{equation}
This representation assumes that atomic densities are centered on the atomic positions $\mathbf R_a$. While this approximation may not be exact for strongly covalent bonds within small fragments, it provides a practical representation of fragment charge distributions.

\paragraph{Atomic decomposition of the electronic density}

In the support-function representation employed in BigDFT, the electronic density is written as
\begin{equation}
\rho(\mathbf r)=\sum_{\alpha\beta}\phi_\alpha(\mathbf r)\,K_{\alpha\beta}\,\phi_\beta(\mathbf r),
\end{equation}
where $\phi_\alpha(\mathbf r)$ are the localized support functions and $K_{\alpha\beta}$ is the density kernel. Each support function is associated with an atomic center. Denoting by $\mathcal A_a$ the subset of support functions centered on atom $a$, the atomic projector is defined as
\begin{equation}
\hat W_a=\sum_{\alpha\in\mathcal A_a} |\phi_\alpha\rangle \langle \phi^\alpha|,
\end{equation}
with $\{|\phi^\alpha\rangle\}$ the dual support-function basis. The density matrix projected onto atom $a$ is then
\begin{equation}
\hat F_a=\hat W_a\hat F,
\end{equation}
and the corresponding atomic contribution to the electronic density reads
\begin{equation}
\rho_a(\mathbf r)=\langle \mathbf r|\hat F_a|\mathbf r\rangle
=\sum_{\alpha\in\mathcal A_a}\sum_\beta
\phi_\alpha(\mathbf r)\,K_{\alpha\beta}\,\phi_\beta(\mathbf r).
\end{equation}
Introducing local coordinates with respect to the atomic center,
\begin{equation}
\mathbf x=\mathbf r-\mathbf R_a,
\end{equation}
the atomic multipole moments are defined from $\rho_a(\mathbf R_a+\mathbf x)$. In particular, the monopole, dipole and quadrupole moments are given by
\begin{align}
Q_0^{(a)} &= \int d\mathbf x\, \rho_a(\mathbf R_a+\mathbf x),\\
p_i^{(a)} &= \int d\mathbf x\, x_i\,\rho_a(\mathbf R_a+\mathbf x),\\
Q_{2,ij}^{(a)} &= \int d\mathbf x\, x_i x_j\, \rho_a(\mathbf R_a+\mathbf x).
\end{align}
These quantities provide the atomic contributions entering the multipole expansion of the fragment electrostatic potential.
When atoms are regarded as elementary fragments, the monopole associated with atom $a$ is given by the zeroth moment of the atom-projected density,
\begin{equation}
Q_0^{(a)}=\int d\mathbf r\, \rho_a(\mathbf r).
\end{equation}
Using the definition of the atom-projected density matrix,
\begin{equation}
\hat F_a=\hat W_a \hat F,
\end{equation}
one obtains
\begin{equation}
Q_0^{(a)}=\mathrm{tr}(\hat F_a)=n_a,
\end{equation}
where $n_a$ is the electronic population of atom $a$. If the multipole expansion is written in terms of the electronic charge density rather than the electron-number density, the corresponding monopole is instead $Q_0^{(a)}=-n_a$.

\paragraph{Multipole expansion of fragment densities}

At sufficiently large distances from the atomic positions, the atomic densities can be approximated by a multipole expansion,
\begin{equation}
\rho_a(\mathbf r-\mathbf R_a)
\simeq
Q_0^{(a)}\delta(\mathbf r-\mathbf R_a)
+
\mathbf p_a \cdot \nabla \delta(\mathbf r-\mathbf R_a)
+
Q^{(a)}_{2,ij}
\partial_i \partial_j \delta(\mathbf r-\mathbf R_a)
+\cdots
\end{equation}
where $Q_0^{(a)}$ is the atomic monopole (charge), $\mathbf p_a$ the dipole moment and $Q^{(a)}_{2,ij}$ the quadrupole tensor. This expansion leads to the electrostatic potential generated by fragment $\mathcal F$,
\begin{equation}
V_{\mathcal F}(\mathbf r)
=
\sum_{a\in \mathcal F}
\left[
\frac{Q_0^{(a)}}{|\mathbf r-\mathbf R_a|}
+
\frac{\mathbf p_a\cdot(\mathbf r-\mathbf R_a)}
{|\mathbf r-\mathbf R_a|^3}
+
\frac{Q^{(a)}_{2,ij}(\mathbf r-\mathbf R_a)_i(\mathbf r-\mathbf R_a)_j}
{2|\mathbf r-\mathbf R_a|^5}
+\cdots
\right].
\end{equation}

\paragraph{Fragment electrostatic interaction}

The electrostatic interaction between two fragments can therefore be written as
\begin{equation}
2E_{FG}^{elst}
=
\int d\mathbf r
\, V_F(\mathbf r)\rho_G(\mathbf r).
\end{equation}
Using the multipole expansion, this interaction becomes a sum over atomic contributions,
\begin{equation}
2E_{FG}^{elst}
\simeq
\sum_{a\in F}\sum_{b\in G}
\left[
\frac{Q_0^{(a)}Q_0^{(b)}}{R_{ab}}
+
\frac{(Q_0^{(a)}\mathbf p_b-Q_0^{(b)}\mathbf p_a)\cdot \mathbf R_{ab}
+\mathbf p_a\cdot\mathbf p_b}{R_{ab}^3}
-
3\frac{(\mathbf p_a\cdot\mathbf R_{ab})(\mathbf p_b\cdot\mathbf R_{ab})}{R_{ab}^5}
+\cdots
\right],
\end{equation}
where $\mathbf R_{ab}=\mathbf R_a-\mathbf R_b$.

In practical applications, the electronic fragment density is represented through the multipole expansion described above, while the ionic contribution associated with the atoms of the fragment is treated at the monopole level with screened nuclear charges $Z^{(a)}$. The resulting electrostatic contribution becomes
\begin{equation}
\int d\mathbf r d\mathbf r'
\frac{
\rho_{el}^F(\mathbf r)\rho_{el}^G(\mathbf r')
+
\rho_{ion}^F(\mathbf r)\rho_{ion}^G(\mathbf r')
}{|\mathbf r-\mathbf r'|}
=
\sum_{a\in F}\sum_{b\in G}
\left[
\frac{Z^{(a)}Z^{(b)}+Q_0^{(a)}Q_0^{(b)}}{R_{ab}}
+
\frac{(Q_0^{(a)}\mathbf p_b-Q_0^{(b)}\mathbf p_a)\cdot \mathbf R_{ab}}
{R_{ab}^3}
+\cdots
\right].
\end{equation}
The fragment–fragment electrostatic interaction energy corresponds to one half of this quantity.

\providecommand{\latin}[1]{#1}
\makeatletter
\providecommand{\doi}
  {\begingroup\let\do\@makeother\dospecials
  \catcode`\{=1 \catcode`\}=2 \doi@aux}
\providecommand{\doi@aux}[1]{\endgroup\texttt{#1}}
\makeatother
\providecommand*\mcitethebibliography{\thebibliography}
\csname @ifundefined\endcsname{endmcitethebibliography}  {\let\endmcitethebibliography\endthebibliography}{}

\end{document}


\maketitle

\section{Additional Heatmaps}

\begin{figure}[htbp]
\centering
\includegraphics[width=1.\textwidth]{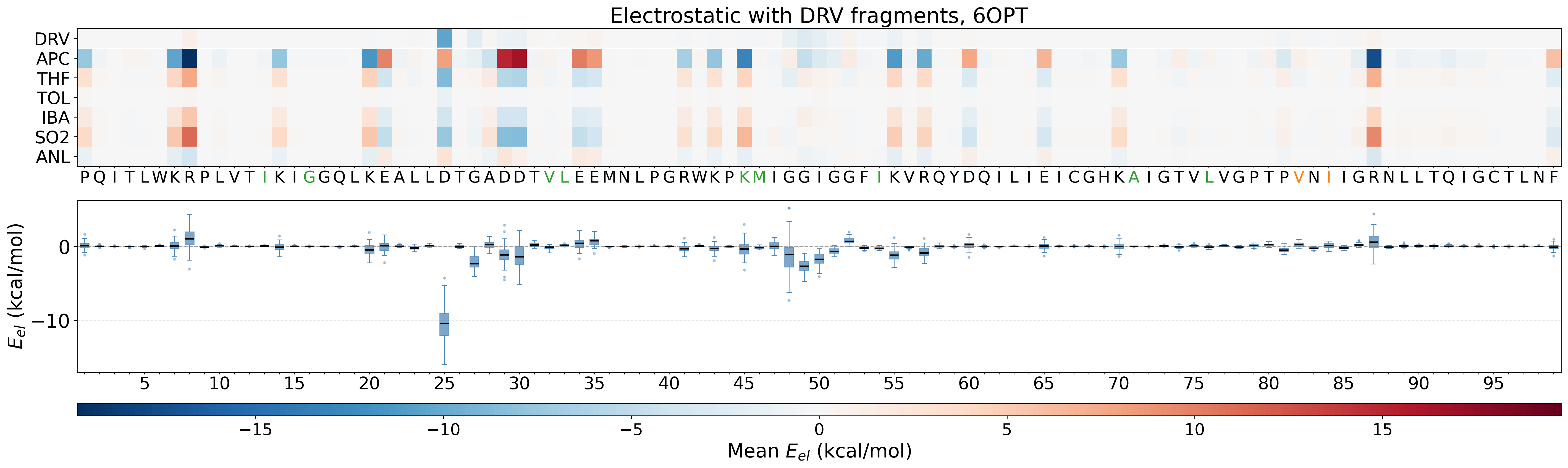}\\
\includegraphics[width=1.\textwidth]{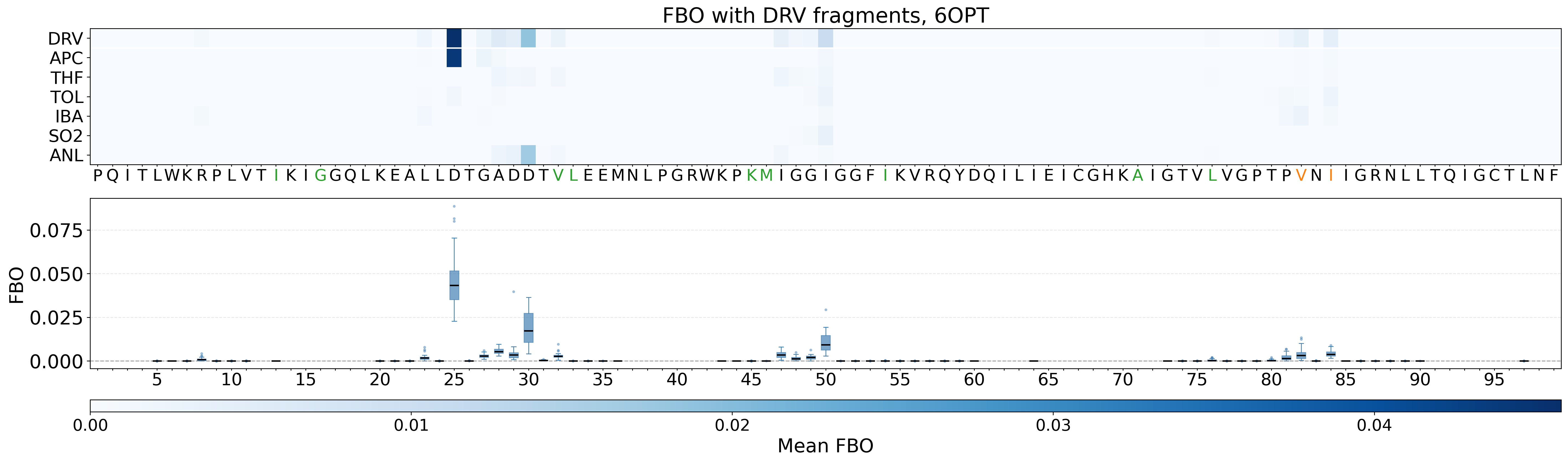}
\caption{Residue-resolved $E_{\mathrm{el}}$ energy (top) and FBO (bottom) between DRV and PR in the 2 mutations 6OPT complex. The average interactions are resolved at a per DRV fragment level (heatmap); we also plot the distribution of interactions for the full DRV molecule. Residues mutated in 6OPT and 6OPZ are written in orange and green, respectively.}
\label{fig:heatmap_6opt}
\end{figure}

\begin{figure}[htbp]
\centering
\includegraphics[width=1.\textwidth]{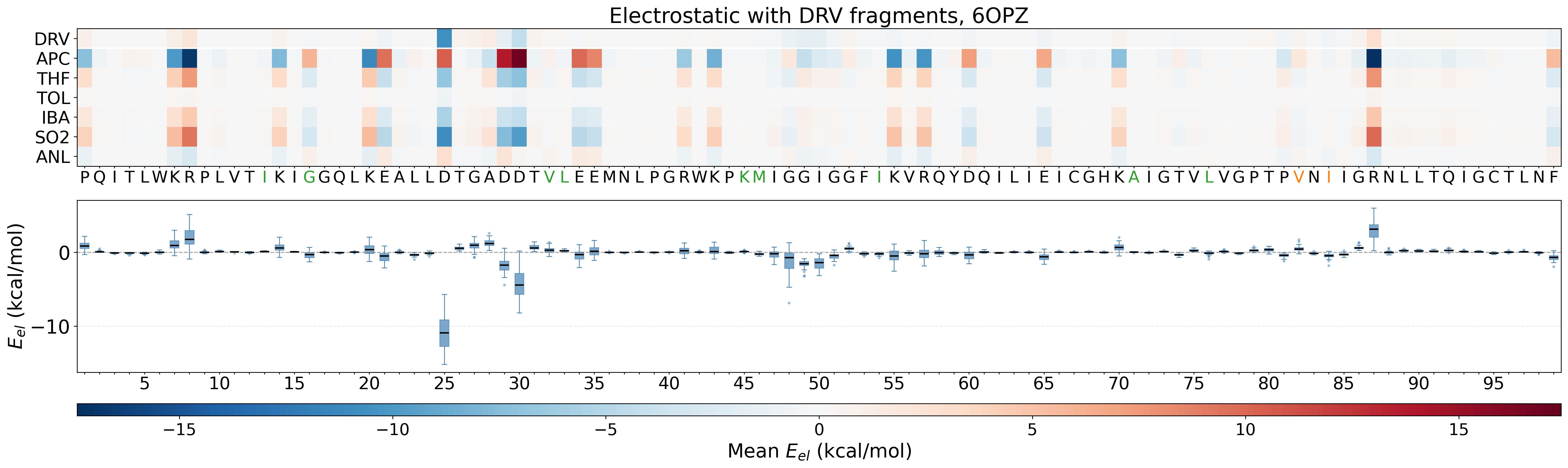}\\
\includegraphics[width=1.\textwidth]{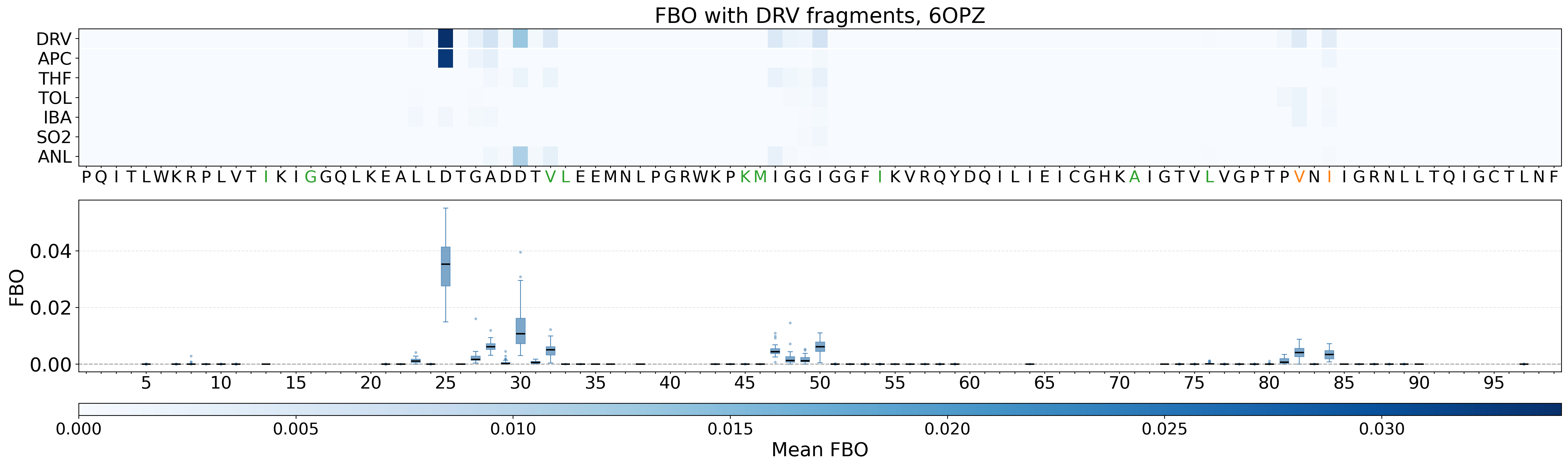}
\caption{Residue-resolved $E_{\mathrm{el}}$ energy (top) and FBO (bottom) between DRV and PR in the 11 mutations 6OPZ complex. The average interactions are resolved at a per DRV fragment level (heatmap); we also plot the distribution of interactions for the full DRV molecule. Residues mutated in 6OPT and 6OPZ are written in orange and green, respectively.}
\label{fig:heatmap_6opz}
\end{figure}

Additional heat maps showing the electrostatic and fragment bond order interactions for the 2 mutation (6OPT) and 11 mutation (6OPZ) variants are plotted in Fig.~\ref{fig:heatmap_6opt} and Fig.~\ref{fig:heatmap_6opz}, respectively.

\section{Pairwise Interaction}

\begin{figure}
    \centering
    \includegraphics[width=1\textwidth]{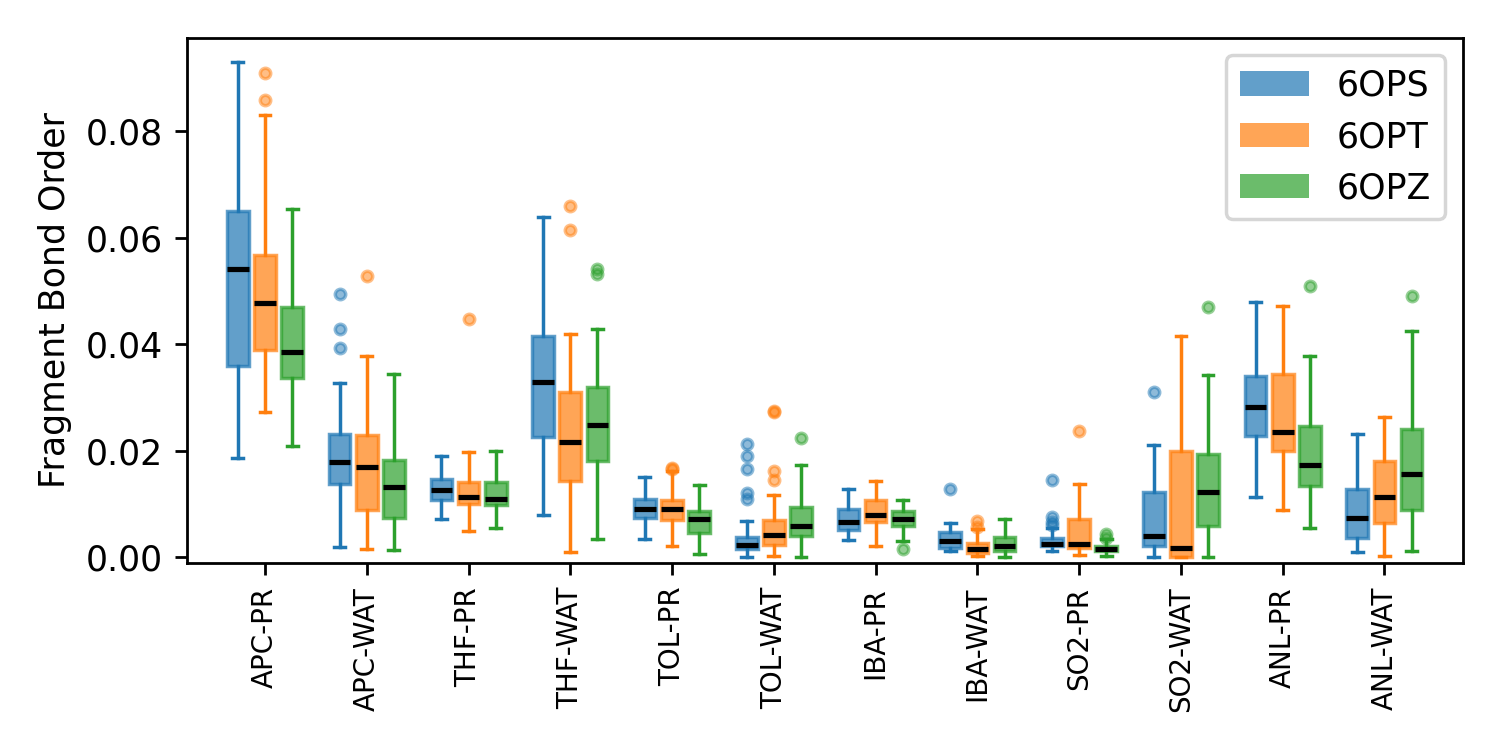}
    \caption{Hotspots of the Darunavir moiety as defined by the FBO measure between the total PR and solution (WAT). We plot the distribution of FBO values for all fragments that have a median FBO above $0.001$.} 
    \label{fig:drv_bond_order}
\end{figure}

To further describe the interaction pattern, we plot in Fig.~\ref{fig:drv_bond_order} the FBO between fragments of DRV and either PR or the water molecules. The main signature of the mutations is the progressive weakening of APC with PR. With ANL, there is a small weakening of the median interaction for 6OPT, but the decrease of the full distribution occurs only after the full set of 6OPZ mutations. Weaker interactions with PR lead to increased opportunities for DRV to bind to water at the TOL and ANL sites. For all three variants, there are consistent interactions between water and APC, THF, and SO2, emphasizing the importance of including explicit water in the simulations.

\begin{figure}
    \centering
    \includegraphics[width=1\textwidth]{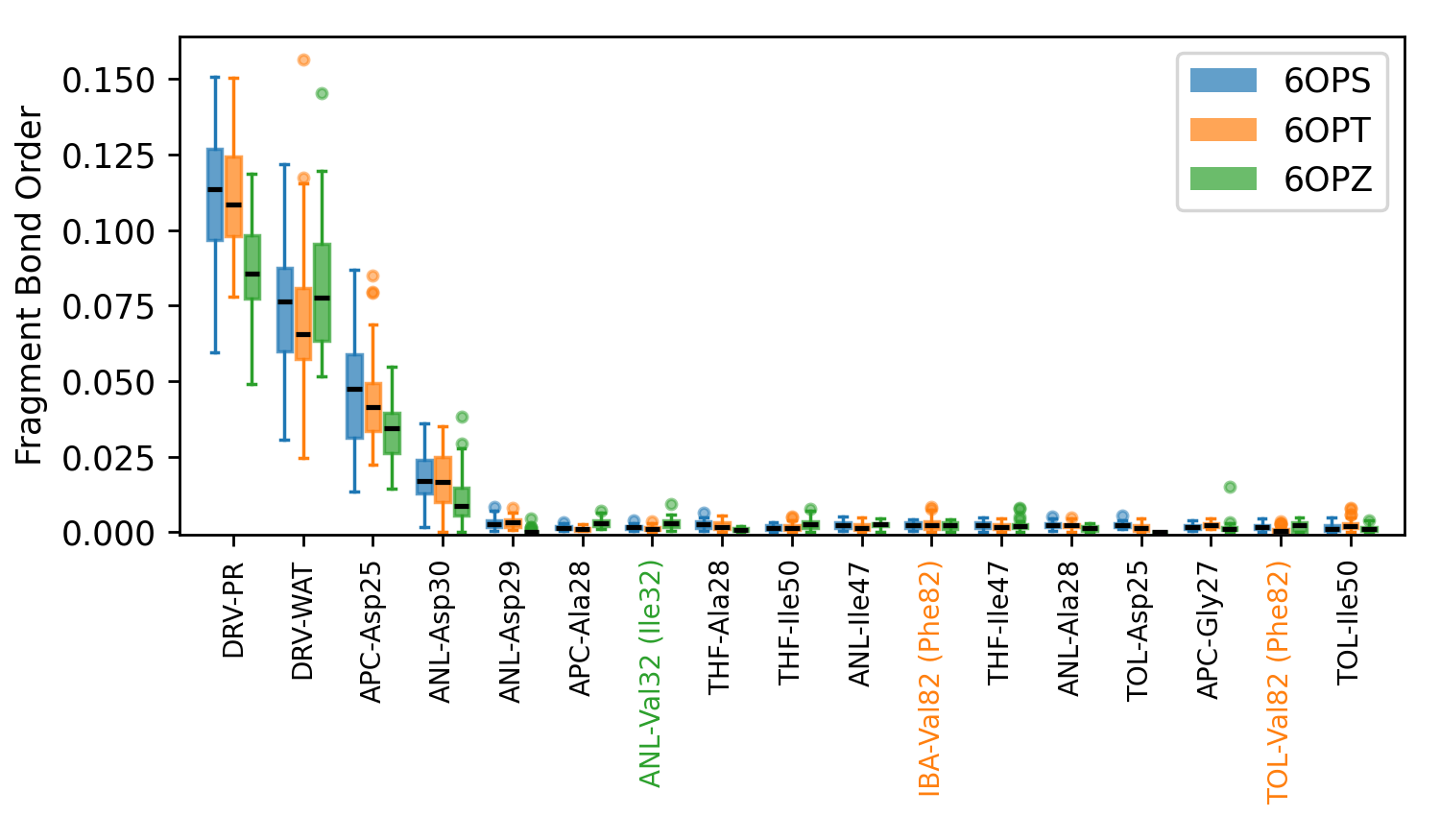} 
    \caption{FBO interactions between amino acid and DRV fragments sorted by median strength. Residues mutated in 6OPT and 6OPZ are written in orange and green, respectively.}
    \label{fig:interaction_pair}
\end{figure}

The specific FBO pairs of protein residues and DRV fragments are plotted in Fig.~\ref{fig:interaction_pair}. There is a notable reduction in interaction between APC and Asp25 as soon as the first two mutations are introduced. An advantage of the selected fragmentation is that we can focus on the interaction between Asp25 and the peptide mimicking portion of DRV: the reduced affinity is a signal for the reduced catalytic activity that the virus must trade for greater drug resistance. In the 6OPZ case, the most significant weakening is between ANL and Asp30.

\begin{figure}
    \centering
    \includegraphics[width=1\textwidth]{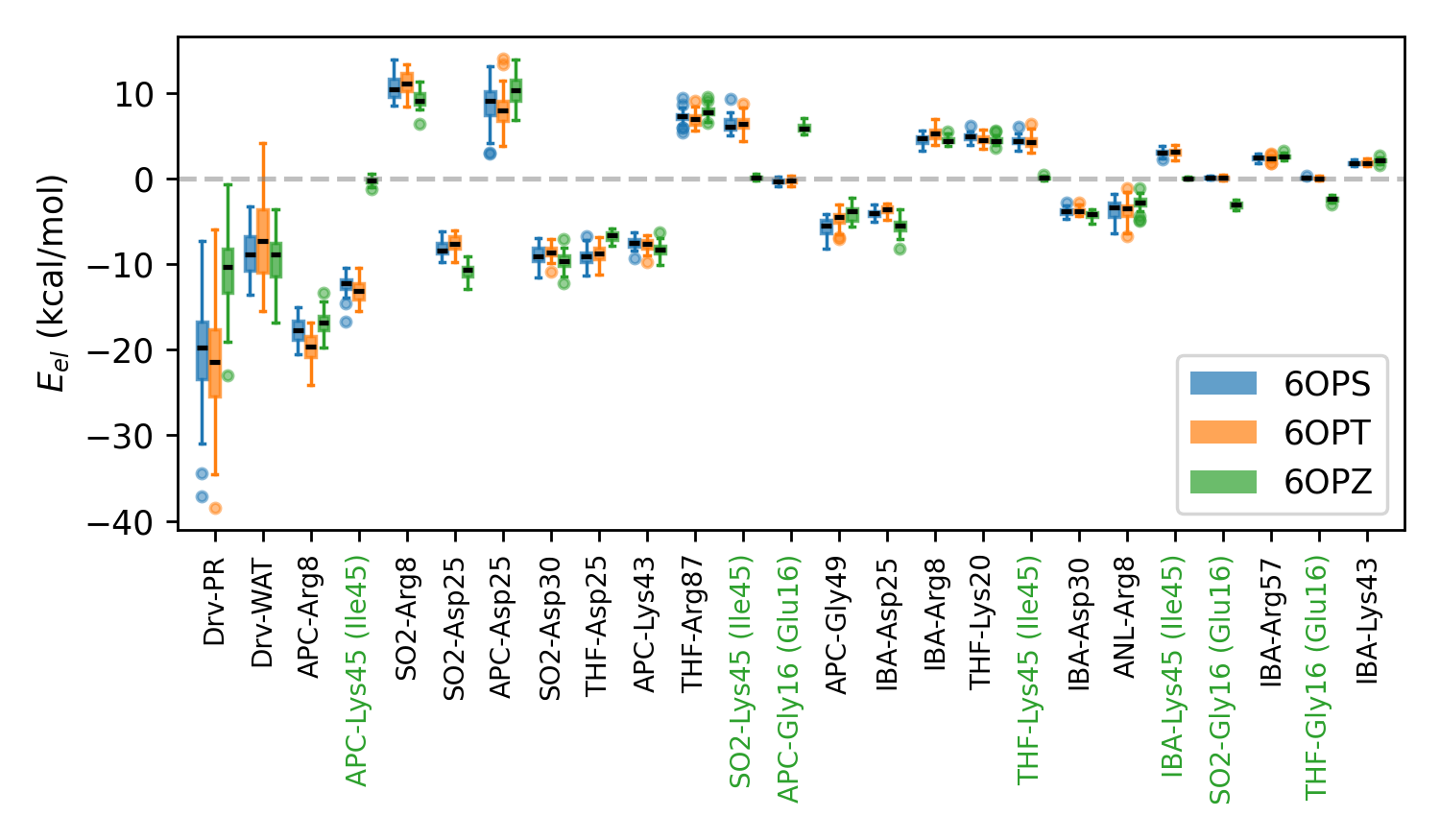} 
    \caption{Electrostatic interactions between amino acid and DRV fragments sorted by median strength. We report only the interactions that are statistically different between the three mutants (narrowed down using a Kolmogorov–Smirnov test with p < 0.001) and have a strength of at least 2 kcal/mol in magnitude. We also plot the distribution of interactions between DRV and PR as well as DRV and the water. Residues mutated in 6OPZ are written in green.}
    \label{fig:electrostatic_pair}
\end{figure}

We also plot the electrostatic interaction pairs between protein residues and DRV fragments in Fig.~\ref{fig:electrostatic_pair}. As a large number of protein residues interact electrostatically, we narrow in on interactions that are statistically different between the variants. The standout cases for modified electrostatic interactions are the mutations G16E and K45I. The introduction of charged mutations in the 6OPZ variant leads to repulsive interactions with the APC group of DRV. However, these interactions are compensated for by increased interactions with SO2 and THF.

